\documentclass[twocolumn]{aastex631}

\usepackage{float}
\usepackage{bm}

\newcommand{\todo}{\ifmmode \text{\color{red}{\Huge(\bullet)}} \else {\color{red}{\Huge$\bullet$}}\fi}
\newcommand{\tido}{\ifmmode {{\color{red}\bullet}} \else {\color{red}{$\bullet$}}\fi}

\newcommand{\E        }[1]{\ifmmode 10^{#1} \else $10^{#1}$\fi}
\newcommand{\tE        }[1]{\ifmmode \times10^{#1} \else $\times10^{#1}$\fi}
\newcommand{\til}{\ifmmode \sim \else $\sim$\fi}
\renewcommand{\~} {\ifmmode \sim \else $\sim$\fi}

\newcommand{\pc}	{\ifmmode \textrm{pc} \else pc\fi}
\newcommand{\kpc}	{\ifmmode \textrm{kpc} \else kpc\fi}
\newcommand{\ld}	{\ifmmode \textrm{l.d.} \else l.d.\fi}
\newcommand{\kms}	{\ifmmode \textrm{km\,s}^{-1} \else km\,s$^{-1}$\fi}
\newcommand{\cc}	{\ifmmode \textrm{cm}^{-3}    \else cm$^{-3}$\fi}
\newcommand{\cmii}	{\ifmmode \textrm{cm}^{-2}    \else cm$^{-2}$\fi}
\newcommand{\ergs}	{\ifmmode \textrm{erg\,s}^{-1} \else erg s$^{-1}$\fi}
\newcommand{\ergcms}	{\ifmmode \textrm{erg\,cm}^{-2}\,\textrm{s}^{-1} \else erg\,cm$^{-2}$\,s$^{-1}$\fi}
\newcommand{\ergcmsA}	{\ifmmode \textrm{erg\,cm}^{-2}\,\textrm{s}^{-1}\,\textrm{\AA}^{-1}
\else erg\,cm$^{-2}$\,s$^{-1}$\,\AA$^{-1}$\fi}
\newcommand{  \ergcmsHz  }{\ifmmode\textrm{erg\,cm}^{-2}\,\textrm{s}^{-1}\,\textrm{Hz}^{-1}
                       \else ergs\,cm$^{-2}$\,s$^{-1}$\,Hz$^{-1}$\fi}
\newcommand{\kev}	{\ifmmode \textrm{keV} \else keV\fi}

\newcommand{\mic}	{\ifmmode \textrm{\mu m} \else $\mu$m\fi}
\newcommand{\vFWHM}	{\ifmmode v_{\mbox{\tiny FWHM}} \else $v_{\mbox{\tiny FWHM}}$\fi}
\newcommand{\vBLR}	{\ifmmode v_{\mbox{\tiny BLR}} \else $v_{\mbox{\tiny BLR}}$\fi}
\newcommand{\sigBLR}	{\ifmmode \sigma_{\mbox{\tiny BLR}} \else $\sigma_{\mbox{\tiny BLR}}$\fi}
\newcommand{\vNLR}	{\ifmmode v_{\mbox{\tiny NLR}} \else $v_{\mbox{\tiny NLR}}$\fi}
\newcommand{\tauBLR}	{\ifmmode \tau_{\mbox{\tiny BLR}} \else $\tau_{\mbox{\tiny BLR}}$\fi}

\newcommand{\Hubble}	{\ifmmode \textrm{km\,s}^{-1}\,\textrm{Mpc}^{-1} \else km\,s$^{-1}$\,Mpc$^{-1}$\fi}
\newcommand{\NDunit}	{\ifmmode \textrm{Mpc}^{-3} \else Mpc$^{-3}$\fi}
\newcommand{\LFunit}	{\ifmmode \textrm{Mpc}^{-3}\,\textrm{mag}^{-1} \else Mpc$^{-3}$\,mag$^{-1}$\fi}
\newcommand{\MFunit}	{\ifmmode \textrm{Mpc}^{-3}\,\textrm{dex}^{-1} \else Mpc$^{-3}$\,dex$^{-1}$\fi}

\newcommand{\Msun}{\ifmmode M_{\odot} \else $M_{\odot}$\fi}
\newcommand{\Lsun}{\ifmmode L_{\odot} \else $L_{\odot}$\fi}
\newcommand{\Zsun}{\ifmmode Z_{\odot} \else $Z_{\odot}$\fi}
\newcommand{\mpyr}{\ifmmode \Msun\,\textrm{yr}^{-1} \else $\Msun\,\textrm{yr}^{-1}$\fi}

\newcommand{\qnote}{\ifmmode q_{0} \else $q_{0}$\fi}
\newcommand{\Hnote}{\ifmmode H_{0} \else $H_{0}$\fi}
\newcommand{\hnote}{\ifmmode h_{0} \else $h_{0}$\fi}
\newcommand{\anote}{\ifmmode a_{0} \else $a_{0}$\fi}
\newcommand{\tnote}{\ifmmode t_{0} \else $t_{0}$\fi}

\newcommand{\gtsim}{\raisebox{-.5ex}{$\;\stackrel{>}{\sim}\;$}}

\newcommand{  \Halpha   }{\ifmmode \textrm{H}\alpha \else H$\alpha$\fi}
\newcommand{  \halpha   }{\Halpha}
\newcommand{  \ha       }{\Halpha}
\newcommand{  \Hbeta    }{\ifmmode \textrm{H}\beta \else H$\beta$\fi}

\newcommand{  \hb       }{\Hbeta}
\newcommand{  \Hgamma   }{\ifmmode \textrm{H}\gamma \else H$\gamma$\fi}
\newcommand{  \Hdelta   }{\ifmmode \textrm{H}\delta \else H$\delta$\fi}
\newcommand{  \Lya      }{\ifmmode \textrm{Ly}\alpha \else Ly$\alpha$\fi}
\newcommand{  \Lyb      }{\ifmmode \textrm{Ly}\beta \else Ly$\beta$\fi}
\newcommand{  \Pa       }{\ifmmode \textrm{P}\alpha \else P$\alpha$\fi}
\newcommand{  \Pb       }{\ifmmode \textrm{P}\beta \else P$\beta$\fi}
\newcommand{  \Bra      }{\ifmmode \textrm{Br}\alpha \else Br$\alpha$\fi}
\newcommand{  \Brg      }{\ifmmode \textrm{Br}\gamma \else Br$\gamma$\fi}
\newcommand{  \hii      }{\ifmmode \textrm{H}\,\textsc{ii} \else H\,\textsc{ii}\fi}
\newcommand{  \hei      }{\ifmmode \textrm{He}\,\textsc{i} \else He\,\textsc{i}\fi}
\newcommand{  \heii     }{\ifmmode \textrm{He}\,\textsc{ii} \else He\,\textsc{ii}\fi}
\newcommand{  \HeIIuv   }{\ifmmode \textrm{He}\,\textsc{ii}\,\lambda1640 \else He\,\textsc{ii}\,$\lambda1640$\fi}
\newcommand{  \HeIIop   }{\ifmmode \textrm{He}\,\textsc{ii}\,\lambda4686 \else He\,\textsc{ii}\,$\lambda4686$\fi}
\newcommand{  \CII	}{\ifmmode \left[\textrm{C}\,\textsc{ii}\right]\,\lambda157.74\,\mu\textrm{m} \else [C\,{\sc ii}]\ $\lambda157.74\,\mu\textrm{m}$\fi}
\newcommand{  \cii	}{\ifmmode \left[\textrm{C}\,\textsc{ii}\right] \else [C\,{\sc ii}]\fi}

\newcommand{  \ciii     }{\ifmmode \textrm{C}\,\textsc{iii}\right] \else C\,\textsc{iii}]\fi}
\newcommand{  \CIII     }{\ifmmode \textrm{C}\,\textsc{iii}\right]\,\lambda1909 \else C\,\textsc{iii}]\,$\lambda1909$\fi}
\newcommand{  \civ      }{\ifmmode \textrm{C}\,\textsc{iv}  \else C\,\textsc{iv}\fi}
\newcommand{  \CIV      }{\ifmmode \textrm{C}\,\textsc{iv}\,\lambda1549 \else C\,\textsc{iv}\,$\lambda1549$\fi}
\newcommand{  \NIIopt   }{\ifmmode \left[\textrm{N}\,\textsc{ii}\right]\,\lambda6584 \else [N\,\textsc{ii}]\,$\lambda6584$\fi}
\newcommand{  \nii      }{\ifmmode \left[\textrm{N}\,\textsc{ii}\right]  \else [N\,\textsc{ii}]\fi}
\newcommand{  \niii     }{\ifmmode \textrm{N}\,\textsc{iii} \else N\,\textsc{iii}\fi}
\newcommand{  \NIII     }{\ifmmode \textrm{N}\,\textsc{iii}\,\lambda4640 \else N\,\textsc{iii}\,$\lambda4640$\fi}
\newcommand{  \niv      }{\ifmmode \textrm{N}\,\textsc{iv}  \else N\,\textsc{iv}\fi}
\newcommand{  \NIVuv    }{\ifmmode \textrm{N}\,\textsc{iv}\,\lambda1486 \else N\,\textsc{iv}\,$\lambda1486$\fi}
\newcommand{  \nv       }{\ifmmode \textrm{N}\,\textsc{v}   \else N\,\textsc{v}\fi}
\newcommand{\oi}{\ifmmode \left[\textrm{O}\,\textsc{i}\right] \else [O\,{\sc i}]\fi}
\newcommand{\OI}{\ifmmode \left[\textrm{O}\,\textsc{i}\right]\,\lambda6300 \else [O\,{\sc i}]$\,\lambda6300$\fi}
\newcommand{\oii}{\ifmmode \left[\textrm{O}\,\textsc{ii}\right] \else [O\,{\sc ii}]\fi}
\newcommand{\OII}{\ifmmode \left[\textrm{O}\,\textsc{ii}\right]\,\lambda3727 \else [O\,{\sc ii}]\,$\lambda3727$\fi}
\newcommand{\oiii}{\ifmmode \textrm{O}\,\textsc{iii} \else O\,{\sc iii}\fi}
\newcommand{\OIII}{\ifmmode \left[\textrm{O}\,\textsc{iii}\right]\,\lambda5007 \else [O\,{\sc iii}]\,$\lambda5007$\fi}
\newcommand{  \OIIIbf   }{\ifmmode \textrm{O}\,\textsc{iii}\,\lambda3133 \else O\,\textsc{iii}\,$\lambda3133$\fi}
\newcommand{  \OIIIuv   }{\ifmmode \textrm{O}\,\textsc{iii}\,\lambda1663 \else O\,\textsc{iii}\,$\lambda1663$\fi}
\newcommand{  \oiv      }{\ifmmode \textrm{O}\,\textsc{iv}  \else O\,\textsc{iv}\fi}
\newcommand{  \OIVuv    }{\ifmmode \textrm{O}\,\textsc{iv}\,\lambda1402  \else O\,\textsc{iv}\,$\lambda1402$\fi}
\newcommand{  \OIVIR    }{\ifmmode \textrm{O}\,\textsc{iv}\,25.9\,\mu \textrm{m} \else O\,\textsc{iv}\,$25.9\,\mu$m\fi}
\newcommand{  \ovi      }{\ifmmode \textrm{O}\,\textsc{vi}   \else O\,\textsc{vi}\fi}
\newcommand{  \Ovi      }{\ifmmode \textrm{O}\,\textsc{vi}\,\lambda1035 \else O\,\textsc{vi}\,$\lambda1035$\fi}
\newcommand{  \nei      }{\ifmmode \textrm{Ne}\,\textsc{i}   \else Ne\,\textsc{i}\fi}
\newcommand{  \neii     }{\ifmmode \textrm{Ne}\,\textsc{ii}  \else Ne\,\textsc{ii}\fi}
\newcommand{  \NeiiIR   }{\ifmmode \textrm{Ne}\,\textsc{ii}\,12.8\,\mu \textrm{m} \else Ne\,\textsc{ii}\,$12.8\,\mu$m\fi}
\newcommand{  \neiii    }{\ifmmode \textrm{Ne}\,\textsc{iii} \else Ne\,\textsc{iii}\fi}
\newcommand{  \neiv     }{\ifmmode \textrm{Ne}\,\textsc{iv}  \else Ne\,\textsc{iv}\fi}
\newcommand{  \nev      }{\ifmmode \textrm{Ne}\,\textsc{v}   \else Ne\,\textsc{v}\fi}
\newcommand{  \NevIR    }{\ifmmode \textrm{Ne}\,\textsc{v}\,24.3\,\mu \textrm{m} \else Ne\,\textsc{v}\,$24.3\,\mu$m\fi}
\newcommand{  \nevi     }{\ifmmode \textrm{Ne}\,\textsc{vi}  \else Ne\,\textsc{vi}\fi}
\newcommand{  \mgi      }{\ifmmode \textrm{Mg}\,\textsc{i} \else Mg\,\textsc{i}\fi}
\newcommand{  \mgii     }{\ifmmode \textrm{Mg}\,\textsc{ii} \else Mg\,\textsc{ii}\fi}
\newcommand{  \MgII     }{\ifmmode \textrm{Mg}\,\textsc{ii}\,\lambda2798 \else Mg\,\textsc{ii}\,$\lambda2798$\fi}
\newcommand{  \sii      }{\ifmmode \textrm{S}\,\textsc{ii} \else S\,\textsc{ii}\fi}
\newcommand{  \siii     }{\ifmmode \textrm{S}\,\textsc{iii} \else S\,\textsc{iii}\fi}
\newcommand{  \siv      }{\ifmmode \textrm{S}\,\textsc{iv} \else S\,\textsc{iv}\fi}
\newcommand{  \sili     }{\ifmmode \textrm{Si}\,\textsc{i}   \else Si\,\textsc{i}\fi}
\newcommand{  \silii    }{\ifmmode \textrm{Si}\,\textsc{ii}  \else Si\,\textsc{ii}\fi}
\newcommand{  \Siliv    }{\ifmmode \textrm{Si}\,\textsc{iv}  \else Si\,\textsc{iv}\fi}
\newcommand{  \SilIVuv  }{\ifmmode \textrm{Si}\,\textsc{iv}\,\lambda1400  \else Si\,\textsc{iv}\,$\lambda1400$\fi}
\newcommand{  \AlIII   }{\ifmmode \textrm{Al}\,\textsc{iii}\,\lambda1857 \else Al\,\textsc{iii}\,$\lambda1857$\fi}
\newcommand{  \Aliii   }{\ifmmode \textrm{Al}\,\textsc{iii} \else Al\,\textsc{iii}\fi}
\newcommand{  \caii     }{\ifmmode \textrm{Ca}\,\textsc{ii} \else Ca\,\textsc{ii}\fi}
\newcommand{  \feii     }{\ifmmode \textrm{Fe}\,\textsc{ii} \else Fe\,\textsc{ii}\fi}
\newcommand{  \feiii    }{\ifmmode \textrm{Fe}\,\textsc{iii} \else Fe\,\textsc{iii}\fi}
\newcommand{  \Kalpha   }{\ifmmode \textrm{K}\alpha \else K$\alpha$\fi}

\newcommand{ \Lhb   }{\ifmmode L_{\hb} \else $L_{\hb}$\fi}
\newcommand{ \Lha   }{\ifmmode L_{\ha} \else $L_{\ha}$\fi}
\newcommand{ \fwhb  }{\ifmmode \textrm{FWHM}\left(\hb\right) \else  FWHM(\hb)\fi}
\newcommand{\sighb  }{\ifmmode \sigma\left(\hb\right) \else $\sigma\left(\hb\right)$\fi}
\newcommand{ \ewhb  }{\ifmmode \textrm{EW}\left(\hb\right) \else EW(\hb)\fi}
\newcommand{ \fwha  }{\ifmmode \textrm{FWHM}\left(\ha\right) \else FWHM(\ha)\fi}
\newcommand{ \ewha  }{\ifmmode \textrm{EW}\left(\ha\right) \else EW(\ha)\fi}
\newcommand{ \Lmg   }{\ifmmode L\left(\mgii\right) \else $L\left(\mgii\right)$\fi}
\newcommand{ \fwmg  }{\ifmmode \textrm{FWHM}\left(\mgii\right) \else FWHM(\mgii)\fi}
\newcommand{ \Lciv  }{\ifmmode L\left(\civ\right) \else $L\left(\civ\right)$\fi}
\newcommand{ \fwciv }{\ifmmode \textrm{FWHM}\left(\civ\right) \else FWHM(\civ)\fi}
\newcommand{ \fwhm  }{\ifmmode \textrm{FWHM} \else FWHM\fi} 
\newcommand{ \voff  }{\ifmmode v_\textrm{off} \else $v_\textrm{off}$\fi} 
\newcommand{ \vmax  }{\ifmmode v_\textrm{max} \else $v_\textrm{max}$\fi} 

\newcommand{ \mumg  }{\ifmmode \mu\left(\mgii\right) \else $\mu\left(\mgii\right)$\fi}
\newcommand{ \fmg   }{\ifmmode f\left(\mgii\right) \else $f\left(\mgii\right)$\fi}
\newcommand{ \muciv }{\ifmmode \mu\left(\civ\right) \else $\mu\left(\civ\right)$\fi}
\newcommand{ \fciv  }{\ifmmode f\left(\civ\right) \else $f\left(\civ\right)$\fi}

\newcommand{  \auvo     }{\ifmmode \alpha_{\nu,\textrm{UVO}} \else $\alpha_{\nu,\textrm{UVO}}$\fi}
\newcommand{  \aox      }{\ifmmode \alpha_{\,\textrm{O\textsc{x}}} \else $\alpha_{\,\textrm{O\textsc{x}}}$\fi}
\newcommand{  \Ledd     }{\ifmmode L_\textrm{Edd} \else $L_\textrm{Edd}$\fi}
\newcommand{  \lamLlam  }{\ifmmode \lambda L_{\lambda} \else $\lambda L_{\lambda}$\fi}
\newcommand{  \lLl      }{\ifmmode \lambda L_{\lambda} \else $\lambda L_{\lambda}$\fi}
\newcommand{  \nuLnu    }{\ifmmode \nu L_{\nu} \else $\nu L_{\nu}$\fi}
\newcommand{  \nLn      }{\ifmmode \nu L_{\nu} \else $\nu L_{\nu}$\fi}
\newcommand{  \Luv      }{\ifmmode L_{1450} \else $L_{1450}$\fi}
\newcommand{  \Lop      }{\ifmmode L_{5100} \else $L_{5100}$\fi}
\newcommand{  \lLop     }{\ifmmode \log\left(\Lop/\ergs\right) \else $\log\left(\Lop/\ergs\right)$\fi}
\newcommand{  \Lthree   }{\ifmmode L_{3000} \else $L_{3000}$\fi}
\newcommand{  \lLthree  }{\ifmmode \log\left(\Lthree/\ergs\right) \else $\log\left(\Lthree/\ergs\right)$\fi}
\newcommand{  \Lsix      }{\ifmmode L_{6200} \else $L_{6200}$\fi}
\newcommand{  \lLisx     }{\ifmmode \log\left(\Lop/\ergs\right) \else $\log\left(\Lop/\ergs\right)$\fi}
\newcommand{  \Lxray    }{\ifmmode L_\textrm{X} \else $L_\textrm{X}$\fi}

\newcommand{  \Lhard    }{\ifmmode L_\textrm{2-10} \else $L_\textrm{2-10}$\fi}
\newcommand{  \Lsoft    }{\ifmmode L_\textrm{0.5-2} \else $L_\textrm{0.5-2}$\fi}

\newcommand{\Fthree}{\ifmmode F_{3000} \else $F_{3000}$\fi}
\newcommand{\fuv}{\ifmmode f_{\lambda}\left(1450\textrm{\AA}\right) \else $f_{\lambda}\left(1450 \textrm{\AA}\right)$\fi}
\newcommand{\fthree}{\ifmmode f_{\lambda}\left(3000\textrm{\AA}\right) \else $f_{\lambda}\left(3000\textrm{\AA}\right)$\fi}
\newcommand{\fH}{\ifmmode f_{\lambda}\left(1.65\micron\right) \else
$f_{\lambda}\left(1.65\micron\right)$\fi}

\newcommand{\fbol}{\ifmmode f_\textrm{bol} \else $f_\textrm{bol}$\fi}
\newcommand{\fbolwv}{\ifmmode f_\textrm{bol}\left(\lambda\right) \else $f_\textrm{bol}\left(\lambda\right)$\fi}
\newcommand{\fbolopt}{\ifmmode f_\textrm{bol}\left(5100\textrm{\AA}\right) \else $f_\textrm{bol}\left(5100\textrm{\AA}\right)$\fi}
\newcommand{\fbolthree}{\ifmmode f_\textrm{bol}\left(3000\textrm{\AA}\right) \else $f_\textrm{bol}\left(3000\textrm{\AA}\right)$\fi}
\newcommand{\fboluv}{\ifmmode f_\textrm{bol}\left(1450\textrm{\AA}\right) \else $f_\textrm{bol}\left(1450\textrm{\AA}\right)$\fi}

\newcommand{\fbolbat}{\ifmmode f_\textrm{bol}\left(14-150\,\kev\right) \else $f_\textrm{bol}\left(14-150\,\kev\right)$\fi}
\newcommand{\fbolhard}{\ifmmode f_\textrm{bol}\left(2-10\,\kev\right) \else $f_\textrm{bol}\left(2-10\,\kev\right)$\fi}

\newcommand{\fobs}{\ifmmode f_\textrm{obs} \else $f_\textrm{obs}$\fi}

\newcommand{\mbh}{\ifmmode M_{\rm BH} \else $M_{\rm BH}$\fi}
\newcommand{  \lmbh     }{\ifmmode \log\left(\mbh/\Msun\right) \else $\log\left(\mbh/\Msun\right)$\fi} 
\newcommand{  \lledd    }{\ifmmode L/L_\textrm{Edd} \else $L/L_\textrm{Edd}$\fi}
\newcommand{  \mmedd    }{\ifmmode \dot{m}/\dot{m}_\textrm{\,Edd} \else $\dot{m}/\dot{m}_\textrm{\,Edd}$\fi}
\newcommand{  \Lbol     }{\ifmmode L_\textrm{bol} \else $L_\textrm{bol}$\fi}
\newcommand{  \lbol     }{\ifmmode L_\textrm{bol} \else $L_\textrm{bol}$\fi}
\newcommand{  \lLbol    }{\ifmmode \log\left(\Lbol/\ergs\right) \else $\log\left(\Lbol/\ergs\right)$\fi} 
\newcommand{  \Lagn     }{\ifmmode L_\textrm{AGN} \else $L_\textrm{AGN}$\fi}
\newcommand{  \lagn     }{\ifmmode L_\textrm{AGN} \else $L_\textrm{AGN}$\fi}

\newcommand{  \tgrow     }{\ifmmode t_\textrm{growth} \else $t_\textrm{growth}$\fi}
\newcommand{  \tAD     }{\ifmmode t_\textrm{acc} \else $t_\textrm{acc}$\fi}
\newcommand{  \tacc    }{\ifmmode t_\textrm{acc} \else $t_\textrm{acc}$\fi}
\newcommand{  \tUni      }{\ifmmode t_\textrm{Universe} \else $t_\textrm{Universe}$\fi}

\newcommand{  \Mdotin	}{\ifmmode \dot{M}_\textrm{infall} \else $\dot{M}_\textrm{infall}$\fi}
\newcommand{  \Mdotbh	}{\ifmmode \dot{M}_\textrm{BH} \else $\dot{M}_\textrm{BH}$\fi}
\newcommand{  \Mdotad	}{\ifmmode \dot{M}_\textrm{AD} \else $\dot{M}_\textrm{AD}$\fi}
\newcommand{  \Mdotacc	}{\ifmmode \dot{M}_\textrm{acc} \else $\dot{M}_\textrm{acc}$\fi}
\newcommand{  \Mdotthin	}{\ifmmode \dot{M}_\textrm{thin} \else $\dot{M}_\textrm{thin}$\fi}
\newcommand{  \Mdotdisk	}{\ifmmode \dot{M}_\textrm{disk} \else $\dot{M}_\textrm{disk}$\fi}

\newcommand{  \Mindot	}{\ifmmode \dot{M}_\textrm{infall} \else $\dot{M}_\textrm{infall}$\fi}
\newcommand{  \Mbhdot	}{\ifmmode \dot{M}_\textrm{BH} \else $\dot{M}_\textrm{BH}$\fi}
\newcommand{  \Maddot	}{\ifmmode \dot{M}_\textrm{AD} \else $\dot{M}_\textrm{AD}$\fi}
\newcommand{  \Maccdot	}{\ifmmode \dot{M}_\textrm{acc} \else $\dot{M}_\textrm{acc}$\fi}
\newcommand{  \Mthdot	}{\ifmmode \dot{M}_\textrm{thin} \else $\dot{M}_\textrm{thin}$\fi}
\newcommand{  \Mdsdot	}{\ifmmode \dot{M}_\textrm{disk} \else $\dot{M}_\textrm{disk}$\fi}

\newcommand{  \as	}{\ifmmode a_\textrm{*} \else $a_\textrm{*}$\fi}
\newcommand{  \avec	}{\ifmmode \vec{a}_\textrm{*} \else $\vec{a}_\textrm{*}$\fi}
\newcommand{  \re	}{\ifmmode \eta      	 \else $\eta$\fi}
\newcommand{  \RISCO	}{\ifmmode R_\textrm{ISCO}  \else $R_\textrm{ISCO}$\fi}

\newcommand{  \mseed    }{\ifmmode M_\textrm{seed} \else $M_\textrm{seed}$\fi}
\newcommand{  \mbul     }{\ifmmode M_\textrm{bulge} \else $M_\textrm{bulge}$\fi} 
\newcommand{  \mstar    }{\ifmmode M_{*} \else $M_{*}$\fi} 
\newcommand{  \mgal     }{\ifmmode M_{*} \else $M_{*}$\fi} 
\newcommand{  \mhost    }{\ifmmode M_\textrm{host} \else $M_\textrm{host}$\fi}
\newcommand{  \mmsmall  }{\ifmmode M_\textrm{BH}/M_{*} \else $M_\textrm{BH}/M_{*}$\fi}
\newcommand{  \mmlarge  }{\ifmmode M_{*}/M_\textrm{BH} \else $M_{*}/M_\textrm{BH}$\fi}

\newcommand{  \mmdotlarge}{\ifmmode \dot{M}_*/\Mbhdot \else $\dot{M}_*/\Mbhdot$\fi}
\newcommand{  \mmdotsmall}{\ifmmode \Mbhdot/\dot{M}_* \else $\Mbhdot/\dot{M}_*$\fi}

\newcommand{  \mmwp     }{\ifmmode \left(M_{*}/M_\textrm{BH}\right) \else $\left(M_{*}/M_\textrm{BH}\right)$\fi}
\newcommand{  \ml       }{\ifmmode M_{*}/L_{*} \else $M_{*}/L_{*}$\fi}
\newcommand{  \mlwp     }{\ifmmode \left(M_{*}/L\right) \else $\left(M_{*}/L\right)$\fi}
\newcommand{  \mlk      }{\ifmmode \left(M_{*}/L_{K}\right) \else $\left(M_{*}/L_{K}\right)$\fi}
\newcommand{  \sigs     }{\ifmmode \sigma_{*} \else $\sigma_{*}$\fi}
\newcommand{  \Reff     }{\ifmmode R_\textrm{e} \else $R_\textrm{e}$\fi}
\newcommand{  \Rvir     }{\ifmmode R_\textrm{vir} \else $R_\textrm{vir}$\fi}
\newcommand{  \Rtwo     }{\ifmmode R_{200} \else $R_{200}$\fi}
\newcommand{  \Rfive    }{\ifmmode R_{500} \else $R_{500}$\fi}
\newcommand{  \Rgrp     }{\ifmmode R_\textrm{grp} \else $R_\textrm{grp}$\fi}
\newcommand{  \nser     }{\ifmmode n_\textrm{s} \else $n_\textrm{s}$\fi}
\newcommand{  \LSF      }{\ifmmode L_\textrm{SF}  \else $L_\textrm{SF}$\fi}
\newcommand{  \LFIR     }{\ifmmode L_\textrm{FIR} \else $L_\textrm{FIR}$\fi}
\newcommand{  \Lfir     }{\ifmmode L_\textrm{FIR} \else $L_\textrm{FIR}$\fi}
\newcommand{  \LTIR     }{\ifmmode L_\textrm{TIR} \else $L_\textrm{TIR}$\fi}
\newcommand{  \Ltir     }{\ifmmode L_\textrm{TIR} \else $L_\textrm{TIR}$\fi}

\newcommand{  \mdyn     }{\ifmmode M_\textrm{dyn} \else $M_\textrm{dyn}$\fi} 
\newcommand{  \mgas     }{\ifmmode M_\textrm{gas} \else $M_\textrm{gas}$\fi} 
\newcommand{  \mh       }{\ifmmode M_\textrm{h} \else $M_\textrm{h}$\fi}
\newcommand{  \mhalo    }{\ifmmode M_\textrm{halo} \else $M_\textrm{halo}$\fi}
\newcommand{  \sfr      }{\ifmmode \textrm{SFR} \else SFR\fi}

\newcommand{ \Lcii     }{\ifmmode L_{\cii} \else $L_{\cii}$\fi}
\newcommand{ \fwcii  }{\ifmmode \textrm{FWHM}\cii \else FWHM\cii\fi}

\newcommand{\Rin}{\ifmmode R_\textrm{in} \else $R_\textrm{in}$\fi}
\newcommand{\Rout}{\ifmmode R_\textrm{out} \else $R_\textrm{out}$\fi}
\newcommand{\RBLR}{\ifmmode R_\textrm{BLR} \else $R_\textrm{BLR}$\fi}
\newcommand{\RNLR}{\ifmmode R_\textrm{NLR} \else $R_\textrm{NLR}$\fi}

\newcommand{  \hst     }  {{\it HST}}

\newcommand{  \swift   }  {{\it Swift}}

\newcommand{\bj}{\ifmmode b_\textrm{J} \else $b_\textrm{J}$\fi}

\newcommand{\iab}{\ifmmode i_\textrm{AB} \else $i_\textrm{AB}$\fi}

\newcommand{\jab}{\ifmmode J_\textrm{AB} \else $J_\textrm{AB}$\fi}
\newcommand{\hab}{\ifmmode H_\textrm{AB} \else $H_\textrm{AB}$\fi}
\newcommand{\kab}{\ifmmode K_\textrm{AB} \else $K_\textrm{AB}$\fi}

\newcommand{\jveg}{\ifmmode J_\textrm{Vega} \else $J_\textrm{Vega}$\fi}
\newcommand{\hveg}{\ifmmode H_\textrm{Vega} \else $H_\textrm{Vega}$\fi}
\newcommand{\kveg}{\ifmmode K_\textrm{Vega} \else $K_\textrm{Vega}$\fi}

\newcommand{  \Chisq    }{\ifmmode \chi^{2} \else $\chi^{2}$}
\newcommand{  \nelec    }{\ifmmode n_{e} \else $n_{e}$\fi}     
\newcommand{  \nh       }{\ifmmode n_\textrm{H} \else $n_\textrm{H}$\fi}     
\newcommand{  \Ncol     }{\ifmmode N_\textrm{col} \else $N_\textrm{col}$\fi} 
\newcommand{  \NH       }{\ifmmode N_\textrm{H} \else $N_\textrm{H}$\fi}     

\def\sun{\hbox{$\odot$}}

\def\arcsec{\hbox{$^{\prime\prime}$}}
\def\ion#1#2{#1$\;${\small\textrm{\@Roman{#2}}}\relax}

\submitjournal{ApJ}

\shorttitle{AT 2021loi}
\shortauthors{Makrygianni et al.}
\graphicspath{{./}{figures/}}

\begin{document}

\title{AT\,2021loi: A Bowen Fluorescence Flare with a Rebrightening Episode, Occurring in a Previously-Known AGN}


\author[0000-0002-7466-4868]{Lydia Makrygianni}
\affiliation{The School of Physics and Astronomy, Tel Aviv University, Tel Aviv 69978, Israel}

\author[0000-0002-3683-7297]{Benny Trakhtenbrot}
\affiliation{The School of Physics and Astronomy, Tel Aviv University, Tel Aviv 69978, Israel}

\author[0000-0001-7090-4898]{Iair Arcavi}
\affiliation{The School of Physics and Astronomy, Tel Aviv University, Tel Aviv 69978, Israel}
\affiliation{CIFAR Azrieli Global Scholars program, CIFAR, Toronto, Canada}

\author[0000-0001-5231-2645]{Claudio Ricci}
\affiliation{N\'ucleo de Astronom\'ia de la Facultad de Ingenier\'ia, Universidad Diego Portales, Av. Ej\'ercito Libertador 441, Santiago, Chile}
\affiliation{Kavli Institute for Astronomy and Astrophysics, Peking University, Beijing 100871, People's Republic of China}

\author[0000-0002-9347-2298]{Marco C. Lam}
\affiliation{The School of Physics and Astronomy, Tel Aviv University, Tel Aviv 69978, Israel}

\author[0000-0002-5936-1156]{Assaf Horesh}
\affiliation{The Racah Institute of Physics, The Hebrew University of Jerusalem, Jerusalem 91904, Israel}

\author[0000-0003-0466-3779]{Itai Sfaradi}
\affiliation{The Racah Institute of Physics, The Hebrew University of Jerusalem, Jerusalem 91904, Israel}

\author[0000-0002-4924-444X]{K. Azalee Bostroem}
\affiliation{Steward Observatory, University of Arizona, 933 North Cherry Avenue, Tucson, AZ 85721-0065, USA}
\affiliation{LSSTC Catalyst Fellow}

\author[0000-0002-0832-2974]{Griffin Hosseinzadeh}
\affiliation{Steward Observatory, University of Arizona, 933 North Cherry Avenue, Tucson, AZ 85721-0065, USA}

\author[0000-0003-4253-656X]{D. Andrew Howell}
\affiliation{Las Cumbres Observatory, 6740 Cortona Drive, Suite 102, Goleta, CA 93117-5575, USA}
\affiliation{Department of Physics, University of California, Santa Barbara, CA 93106-9530, USA}

\author[0000-0002-7472-1279]{Craig Pellegrino}
\affiliation{Las Cumbres Observatory, 6740 Cortona Drive, Suite 102, Goleta, CA 93117-5575, USA}
\affiliation{Department of Physics, University of California, Santa Barbara, CA 93106-9530, USA}

\author{Rob Fender}
\affiliation{Astrophysics, Department of Physics, University of Oxford, Keble Road, Oxford OX1 3RH, UK}
\affiliation{Department of Astronomy, University of Cape Town, Private Bag X3, Rondebosch 7701, South Africa}

\author[0000-0003-3189-9998]{David A. Green}
\affiliation{Astrophysics Group, Cavendish Laboratory, 19 J. J. Thomson Avenue, Cambridge CB3 0HE, UK}

\author[0000-0001-7361-0246]{David R. A. Williams}
\affiliation{Jodrell Bank Centre for Astrophysics, School of Physics and Astronomy, University of Manchester, Manchester M13 9PL, UK}

\author[0000-0002-7735-5796]{Joe Bright}
\affiliation{Astrophysics, Department of Physics, University of Oxford, Keble Road, Oxford OX1 3RH, UK}

\begin{abstract}
AT\,2021loi is an optical-ultraviolet transient located at the center of its host galaxy. Its spectral features identify it as a member of the ``Bowen Fluorescence Flare'' (BFF) class. The first member of this class was considered to be related to a tidal disruption event, but enhanced accretion onto an already active supermassive black hole was suggested as an alternative explanation. AT\,2021loi, having occurred in a previously-known unobscured AGN, strengthens the latter interpretation. Its light curve is similar to those of previous BFFs, showing a rebrightening approximately one year after the main peak (which was not explicitly identified, but might be the case, in all previous BFFs). An emission feature around 4680\AA, seen in the pre-flare spectrum, strengthens by a factor of $\sim$2 around the optical peak of the flare, and is clearly seen as a double peaked feature then, suggesting a blend of \NIII\ with \HeIIop\ as its origin. The appearance of \OIIIbf\ and possible $\niii\,\lambda\lambda4097,4103$ (blended with \Hdelta) during the flare further support a Bowen Fluorescence classification. Here, we present ZTF, ATLAS, Keck, Las Cumbres Observatory, NEOWISE-R, \swift\, AMI and VLA observations of AT\,2021loi, making it one of the best observed BFFs to date. AT\,2021loi thus provides some clarity on the nature of BFFs but also further demonstrates the diversity of nuclear transients.
\end{abstract}


\section{Introduction} \label{sec:intro}

Actively accreting supermassive black holes (SMBHs), commonly referred to as active galactic nuclei (AGN), exhibit variable emission across the electromagnetic spectrum \citep{Fahlman1975, Paterson1994, Clavel1991}, and over a wide range of timescales. This ``normal'' AGN variability is stochastic, and varies from a few percent (over weeks) up to $\sim$20\% (over decades; e.g., \citealt{MacLeod2012a}). Various models have been proposed to explain this stochastic optical/UV variability, motivated by accretion theory and by the observed variability phenomenology. These include accretion disk instabilities, microlensing, X-ray reprocessing of disk thermal emission, and changes in the accretion rate \cite[see, e.g.,][and references therein]{Krolik1991, Hawkins1993, Kawaguchi1998, Pereyra2006, DexterAgol2011,Ruan2014, Caplar2017}. 

Recently, more extreme optical variability has been identified in a number of AGN where the optical/UV brightness varies by a few magnitudes in relatively short timescales ($\sim$few months; e.g., \citealt{Graham2017,Cannizzaro2020,Rumbaugh2018,Graham2020}; see also \citealt{Lawrence2018} and references therein). In addition, there are AGN where the change in brightness is also associated with spectral transitions, i.e., the continuum emission becoming bluer or redder. Rarely, more significant spectral (and related classification) transitions are seen with the appearance or the disappearance of the blue continuum and/or broad line emission that is typical of persistent, broad-line (unobscured) AGN. The latter type of events is commonly referred to as changing-look AGN \cite[e.g.,][]{LaMassa2015, MacLeod2019,Trakhtenbrot2019b,RT2022}\footnote{Here, we refer to changing-look events in terms of their spectral appearence in the optical regime, and not in the X-ray regime.}.
\begin{figure*}
    \centering
    \includegraphics[width=0.95\textwidth]{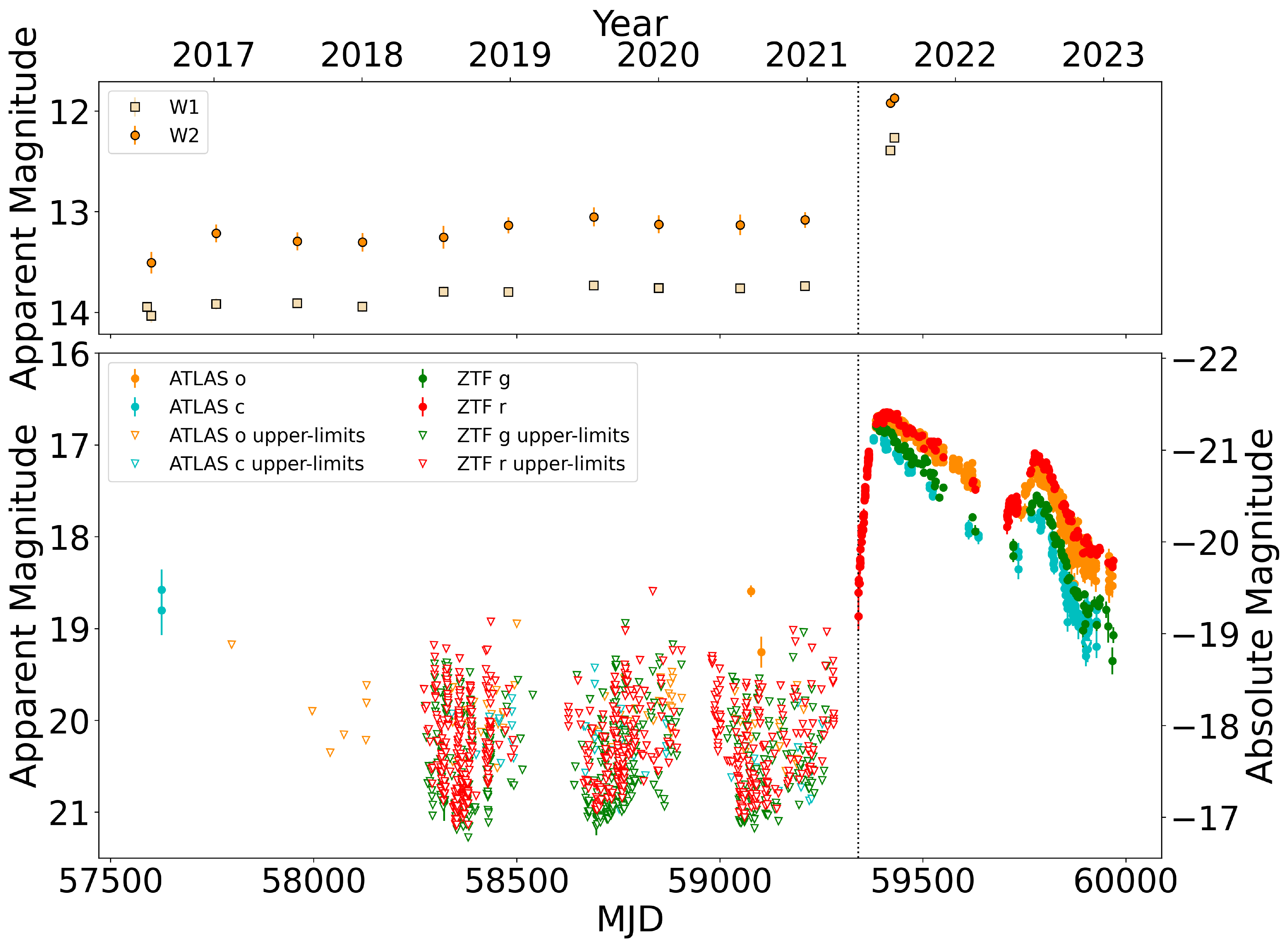}
    \caption{Top: NEOWISE-R $W1$ and $W2$ lightcurves (Vega magnitudes) at the position of AT\,2021loi (not reference subtracted). The reference subtracted lightcurves show no significant variability prior to the optical outburst. The outburst itself is seen in both NEOWISE-R bands. Bottom: Long-term ($\sim4$ years) optical (ZTF $g$ and $r$ bands, ATLAS $c$ and $o$ bands) lightcurves of WISEA J010039.62+394230.3 which shows no significant variability above the non-detection 5$\sigma$ upper limits. A real detection from ATLAS several years ago could correspond to a standard higher amplitude variability event of the unobscured AGN and is much lower than the recent flare. The vertical dotted line in both panels indicates the date of discovery of the optical outburst (2021 May 7, MJD 59341).}
    \label{fig:ltlc}
\end{figure*}

Among the various types of SMBH accretion events that show a large increase in optical/UV flux, \cite{Trakhtenbrot2019a} identified a new class of flares, originally consisting of a brightening event in the ULIRG F01004-2237 \citep{Tadhunter2017}, and the transients AT\,2017bgt \citep{Trakhtenbrot2019a}, and OGLE17aaj \citep{Gromadzki2019}. All of these events were long lasting, unlike other SMBH related transients such as tidal disruption events \cite[TDEs;][]{Rees1988, Gezari2012, Arcavi2014,Gezari2021_rev,VV2020}. During the flare, the spectra of the \cite{Trakhtenbrot2019a} class display emission lines typical of broad-line, unobscured AGN. Importantly, they also exhibit \NIII\ and \OIIIbf\ emission lines, which are not generally seen in AGN \cite[e.g.,][]{VandenBerk2001} but are produced by Bowen Fluorescence (hereafter BF; \citealt{Bowen1928}) in high-velocity (a few 1000s of \kms) and dense gas in the vicinity of the accreting SMBH, as proposed several decades ago by \cite{Netzer1985}. Hereafter, we thus refer to this class of events as ``Bowen Fluorescence Flares'' or BFFs. 

In the BF mechanism, He\,\textsc{ii}\ \Lya{} photons can either escape, ionize neutral H or He, or be absorbed by \oiii\, at 303.693 or 303.799 \AA. These ions are then de-excited through a series of transitions, producing \oiii\, optical lines at 3047, 3133, 3312, 3341, 3444, and 3760 \AA\, as well as a prominent FUV transition at 374.436 \AA. This latter transition can be re-absorbed by ground-state \niii\, which can then produce emission lines at 4097, 4104, 4379, 4634 and 4641 \AA. The presence of strong Bowen lines is thus an indicator of extreme ultraviolet (EUV, down to 100 \AA) radiation, and their observed widths associate them with the broad line region (BLR). An accretion origin could explain both the Bowen lines and perhaps the optical luminosity of these events.

BF lines have been identified in other SMBH-related transients as well. \cite{Blagorodovna2017} and \cite{Leloudas2019} identified BF lines in the spectra of optical TDEs. \cite{Malyali2021} found BF lines in the peculiar nuclear transient AT\,2019avd which was associated with an extreme X-ray flare. \cite{Frederick2021} found BF lines in optical transients occurring in narrow line Seyfert 1 galaxies (NLSy1s) and suggest that three of these events (AT\,2019pev, AT\,2019avd and AT\,2019brs) are BFFs. It is currently not yet clear which properties of the SMBHs, their (variable) accretion flows, and/or their circumnuclear gas environments give rise to transient BF emission.    

Here we report optical, UV, X-ray, and radio observations of AT\,2021loi, a new member of the BFF class. AT\,2021loi is unique as, to our knowledge, it is the only BFF so far to occur in a previously-known broad-line AGN. In addition, it presents the strongest evidence among sources of this class (to date) for a significant rebrightening, roughly one year after its main peak.

This paper is structured as follows: in Sections \ref{discover} and \ref{obs}, we report the discovery and detail the observations, and in Section \ref{analysis}, we present the results of their analysis. In Section \ref{discuss}, we discuss the possible nature of AT\,2021loi and we conclude in Section \ref{concl}. We adopt a flat $\Lambda$CDM cosmology throughout, with $H_0=70\,\rm{\kms Mpc^{-1}}$, $\Omega_M=0.3$ and $\Omega_\Lambda=0.7$. All magnitudes are reported in the AB system \citep{Oke1974}, and all wavelengths are in the rest frame, unless otherwise stated.

\section{Discovery}
\label{discover}

An optical brightening at the center of the $z = 0.083$ active galaxy WISEA J010039.62+394230.3 was discovered in Zwicky Transient Facility (ZTF) public survey data as ZTF20aanxcpf on 2021 May 7 (UT used throughout) by \cite{tns_discovery}, using the ALeRCE broker \citep{Forster2021}. The brightening was reported with an $r$-band reference subtracted\footnote{Throughout the text, the term ``reference'' in the context of photometry or spectroscopy is used to denote the pre-flare emission from the source, which  contains both the stellar light of the host and the pre-flare AGN emission.} discovery magnitude of $18.84\pm0.15$, at R.A. \texttt{22:11:21.93} and Dec \texttt{+39:42:30.31} (J2000). 

The event was classified by \cite{tns_classification} on 2021 June 11 (close to the optical peak at 2021 June 22) as a BFF related to enhanced accretion around an SMBH, based on a spectrum with the Low-Resolution Imaging Spectrograph \citep[LRIS;][]{Oke1995} on Keck I that shows broad Balmer emission lines, consistent with a narrow line Seyfert 1 (NLSy1), but also \HeIIop, \NIII\ and \oiii\ 3133 and 3444 \AA\ emission, not seen in an archival LAMOST spectrum of the source~(see Sec.~3.2).


\section{Observations}
\label{obs}

We collected follow-up observations of AT\,2021loi with various facilities as detailed below. 

\subsection{Photometry}

\subsubsection{Optical, MIR and UV}
\label{sec:phot_data}

\begin{figure*}[h]
    \centering
    \includegraphics[width=0.95\textwidth,height=0.95\textheight]{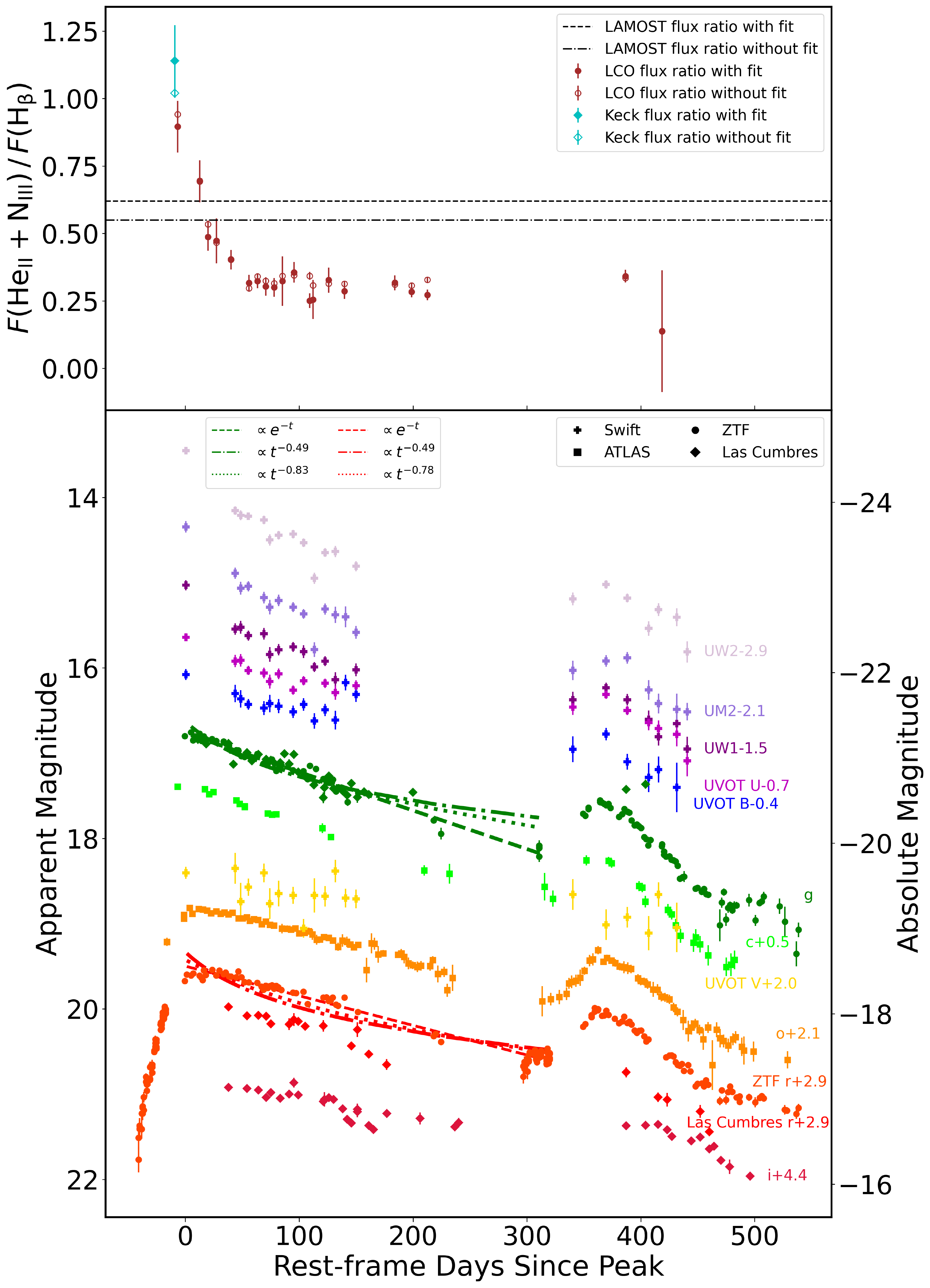}
    \caption{Bottom: Optical and UV lightcurves obtained with ZTF, ATLAS, Las Cumbres and \swift/UVOT. ATLAS photometry is binned in 4-day time spans. \swift\ $U$, $B$ and $V$ magnitudes are not reference subtracted. We present two power-law fits --one with free $t_0$ (dotted lines) and one with $t_0$ fixed to the date of discovery (dashed dotted lines), as well as an exponential fit, to the ZTF $g$ and $r$ band data. The best fit power-laws are shallower than the power law expected for TDEs i.e. $t^{-5/3}$. Top: the evolution of the relative flux of the spectral region that combines \HeIIop\ and \NIII\ to the flux of \hb\, with and without Gaussian modeling of the emission lines.}
    \label{fig:lc}
\end{figure*}

We retrieved PSF-fit photometry of reference subtracted images from the ZTF forced photometry service\footnote{\url{https://ztfweb.ipac.caltech.edu/cgi-bin/requestForcedPhotometry.cgi}} from MJD = 58194 to MJD = 59975, and the  Asteroid Terrestrial-impact Last Alert System \citep[ATLAS;][]{Tonry2018} forced photometry service\footnote{\url{https://fallingstar-data.com/forcedphot/}}, from MJD = 57232 to MJD = 59975 at the position of AT\,2021loi. We find no signs of strong variability in the years preceding the flare (Fig. \ref{fig:ltlc}).

\begin{figure}
    \centering
    \includegraphics[width=0.45\textwidth]{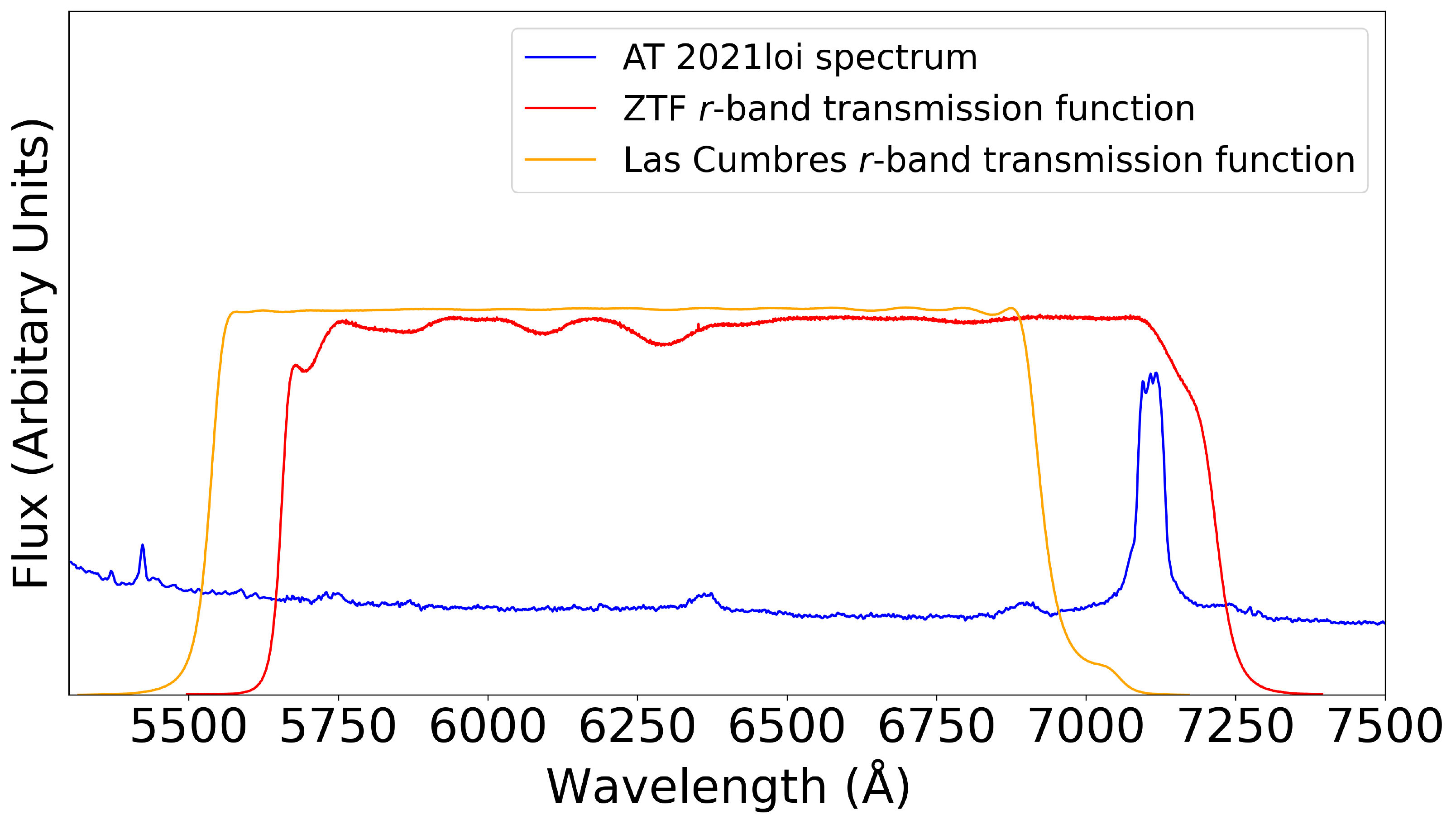}
    \caption{One of the optical spectra of AT\,2021loi plotted along with the ZTF and Las Cumbres $r$ band throughputs. The throughputs have been scaled for visualization purposes. Given the redshift of AT\,2021loi ($z = 0.083$), the ZTF $r$-band includes the \halpha\ line, while the Las Cumbres one does not. This can obviously affect the corresponding flux measurements and account for the discrepancies between the corresponding lightcurves.}
    \label{fig:rvsr}
\end{figure}

We also obtained optical photometry in the $gri$ bands, starting on 2021 July 01, with the Las Cumbres Observatory network of 1\,m telescopes \citep{Brown2013}. As there are no available pre-transient Las Cumbres images of AT\,2021loi, and it has yet to fade, we use Sloan Digital Sky Survey \citep[SDSS;][]{York2000} reference images obtained on 2006 October 6, for all three bands.
We used \textsc{lcogtsnpipe}\footnote{\url{https://github.com/LCOGT/lcogtsnpipe}}\citep[][]{Valenti2016} in order to process the Las Cumbres data. The pipeline is used to generate the Point Spread Function (PSF) for the Las Cumbres images and for the SDSS reference images. After that, the pipeline uses an implementation of the High Order Transform of PSF ANd Template Subtraction \citep[\textsc{hotpants};][]{Becker2015} to perform image subtraction.  We performed aperture photometry at the source position as we find that, for the Las Cumbres data, PSF-fitting produces varying residuals in the image after subtracting the PSF model. There is a noticeable difference between the ZTF and Las Cumbres $r$-band measurements, where the latter appears to be fainter (Figure \ref{fig:lc}). This difference is likely due to the difference in the transmission profiles of the two $r$ filters. 
In Figure \ref{fig:rvsr} we demonstrate this difference using an example optical spectrum of AT\,2021loi along with the two $r$-band filter transmission functions.{\footnote{\url{https://lco.global/observatory/instruments/filters/}}\textsuperscript{,}\footnote{\url{https://github.com/ZwickyTransientFacility/ztf_information/tree/master/filter_transmission}}} While the ZTF filter encompasses the observed broad \halpha\ emission line of AT\,2021loi, the Las Cumbres filter does not, thus explaining why more flux is observed in ZTF $r$ vs. Las Cumbres $r$.

The location of AT\,2021loi was repeatedly observed in the mid-infrared (MIR) regime by the  Wide-Field Infrared Survey Explorer \citep[WISE;][]{Wright2010} and its extensions NEOWISE \citep{Mainzer2011} and the NEOWISE reactivation mission survey, \citep[NEOWISE-R;][]{Mainzer2014}. The measurements were conducted in both the $W1$ and $W2$ bands ($3.4$ and $4.6\,\rm \mu m$, respectively).
We queried the NASA/IPAC Infrared Science Archive\footnote{\url{https://irsa.ipac.caltech.edu/frontpage/}} for MIR detections within 5\arcsec\ from the ZTF-determined position of AT\,2021loi. 
This position was visited roughly twice a year by NEOWISE-R (changes in the scanning pattern may result more observations like in this case where there are the two measurements within a few days in 2016 and 2021). We rebinned the measurements available for each such visit (using a weighted mean) into a single representative measurement per year per band. 
The resulting MIR lightcurve \footnote{MIR magnitudes and colors are given in the Vega-equivalent system} is shown in Figure~\ref{fig:ltlc} (top panel). 
Prior to the optical outburst, the MIR flux shows limited variability of up to a few tenths of magnitude. The first MIR visit to show a significant flux increase ($>1.5$ mag, in both MIR bands), is indeed the first measurement taken after the optical outburst, on 2021 July 22, i.e. ${\sim}70$ days after the initial optical detection of the transient and ${\sim}30$ days after the peak optical emission.

The earliest available WISE-based measurements associated with the the source hosting AT\,2021loi \cite[obtained through the AllWISE catalog;][]{Cutri2014}, using coadded images taken during 2010 December 11 to 2011 January 17, yield a MIR color of $W1-W2 = 0.72$.
The average MIR color during the period 2014 January 18 to 2020 December 31, probed by NEOWISE-R, i.e. the $\sim$7 years prior to the optical outburst  (part of it, $\sim$4.5 years, is shown in Fig.~\ref{fig:ltlc}) is $W1-W2 = 0.66$.
These MIR colors are consistent with the emission being related to an AGN, typically defined at  $W1-W2 \gtrsim 0.7$ \cite[e.g.,][]{Stern2012,Assef2018}.

We initiated near-UV (NUV) follow-up observations (PI: L.\ Makrygianni) with the Neil Gehrels Swift Observatory \citep{Gehrels2004}, Optical/UV Telescope \citep[UVOT;][]{Roming2005}. We obtained twenty-three \swift\ epochs from 2021 June 21 to 2022 October 11 in all six UVOT filters ($uvw2$, $uvm2$, $uvw1$, $U$, $B$ and $V$). UVOT photometry was extracted using the HEASARC pipeline and the standard analysis task \texttt{uvotsource}. We used a circular 7.5\arcsec\ aperture for both the source and the sky extraction regions. We use the same aperture to extract an archival NUV flux measurement from GALEX observations taken on 2006 November 09 for which we find $m_{\rm NUV}=20.05\pm0.14$ mag.

\begin{deluxetable}{ccccc}
\tabletypesize{\footnotesize}
\tablecolumns{5} 
\tablecaption{\label{tab:phot}Photometry of WISEA J010039.62+394230.3.} 
\tablehead{\colhead{\hspace{.33cm}Filter}\hspace{.33cm} & \colhead{\hspace{.33cm} Magnitude}\hspace{.33cm}& \colhead{\hspace{.33cm} Error}\hspace{.33cm}& \colhead{\hspace{.33cm} Source}\hspace{.33cm}&
\colhead{\hspace{.33cm} System}\hspace{.33cm}} 
\startdata
\hline
$FUV$ &  19.86 & 0.21 & GALEX & AB\\
$NUV$ &  20.05 & 0.14 & GALEX & AB\\ 
$u$ &  18.57 & 0.02 & SDSS & AB \\ 
$g$ &  17.91 & 0.01 & SDSS & AB \\ 
$r$ &  17.43 & 0.01 & SDSS & AB \\ 
$i$ &  17.00 & 0.01 & SDSS & AB\\ 
$z$ &  16.82 & 0.01 & SDSS & AB\\ 
$W1$ &  13.64 & 0.03 & WISE & Vega\\ 
$W2$ &  12.92 & 0.03 & WISE & Vega 
\enddata 
\tablecomments{WISEA J010039.62+394230.3, pre-flare photometry without extinction correction applied.}
\label{tab:hostphot}
\end{deluxetable}
Pre-flare photometry of WISEA J010039.62+394230.3 is presented in Table \ref{tab:hostphot}. For AT\,2021loi, all photometry were corrected for Milky Way extinction using $E(B-V)=0.045$ mag \citep{Schlafly2011},\footnote{Retrieved via the NASA/IPAC Extragalactic Database (NED): \url{http://ned.ipac.caltech.edu/}.} the \cite{Cardelli1989} extinction law, and $R_V=3.1$. Our photometry is presented in Figure \ref{fig:ltlc}, in the bottom panel of Figure \ref{fig:lc} and in Table \ref{tab:phot}.

\begin{deluxetable}{ccccc}
\tabletypesize{\footnotesize}
\tablecolumns{5} 
\tablecaption{\label{tab:phot}Photometry of AT\,2021loi.} 
\tablehead{\colhead{MJD} & \colhead{\hspace{.33cm}Filter}\hspace{.33cm} & \colhead{\hspace{.33cm} Magnitude}\hspace{.33cm}& \colhead{\hspace{.33cm} Error}\hspace{.33cm}& \colhead{\hspace{.33cm} Source}\hspace{.33cm} } 
\startdata
\hline
59341.49 &$r$ &  18.99 & 0.15 & ZTF \\        
59341.49 &$r$ &  18.73 & 0.14 & ZTF   \\     
59342.49 & $r$ &  18.63 & 0.08 & ZTF    \\    
59342.49 &$r$ &  18.59 & 0.09 & ZTF    \\    
59344.49 &$r$ &  18.45 & 0.08 & ZTF    \\    
59344.49 &$r$ &  18.63 & 0.11 & ZTF     \\   
59345.48 &$r$ &  18.45 & 0.08 & ZTF     \\   
59345.49 &$r$ &  18.36 & 0.08 & ZTF     \\   
59345.49 &$r$ &   18.38 & 0.08 & ZTF  \\
........ & ...&....&...&... \\
59385.39 &$r$ &   16.89 &0.02 &ZTF \\        
59387.44 &$r$ &   16.82 &0.01 &ZTF  \\      
59391.41 &$r$ &   16.83 &0.02 &ZTF   \\     
59393.45 &$r$ &   16.80 &0.01 &ZTF    \\    
59399.39 &$r$ &   16.83 &0.02 &ZTF    \\    
59401.44 &$r$ &  16.86 &0.01 &ZTF      \\  
59403.43 &$r$ &  16.77 &0.01 &ZTF       \\ 
59405.43 &$r$ &  16.82 &0.01 &ZTF  \\
........&...&....&...&...\\
59428.20 &$g$ &   17.64 & 0.01& Las Cumbres \\
59444.40 &$g$ &  17.33 & 0.03 &Las Cumbres \\
59450.10 &$g$ &  17.24 &0.02 &Las Cumbres  \\
59461.20 &$g$ &  17.33 &0.02 &Las Cumbres \\
59469.40 &$g$ &  17.69 &0.01 &Las Cumbres 
\enddata 
\tablecomments{This table is published in its entirety in machine-readable format. A portion is shown here for guidance regarding its form and content.}
\label{tab:phot}
\end{deluxetable} 

\subsubsection{X-ray and Radio}
\label{subsec:xrayradio}

\begin{figure*}
    \centering
    \includegraphics[width=0.49\textwidth]{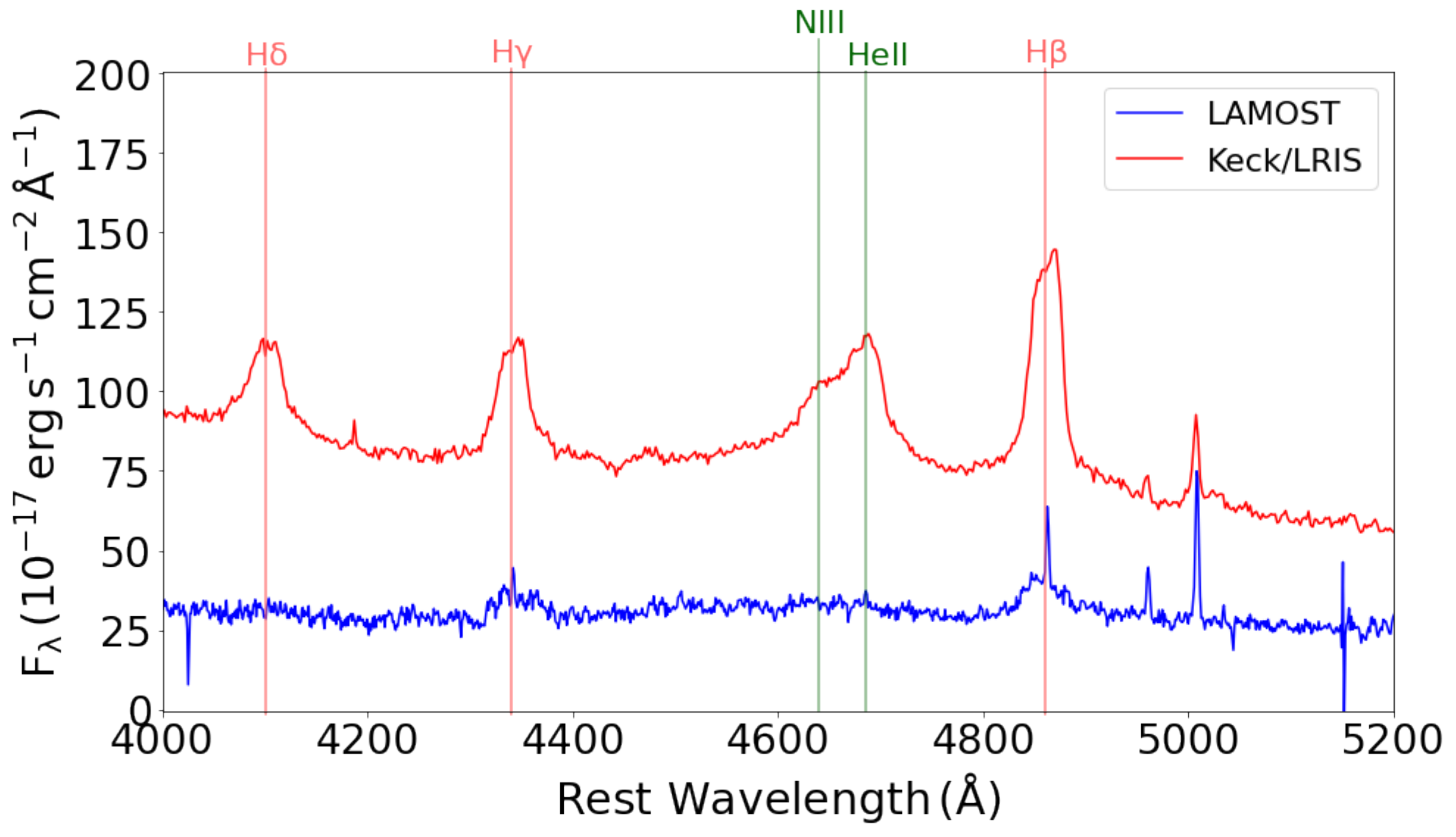}
    \includegraphics[width=0.49\textwidth]{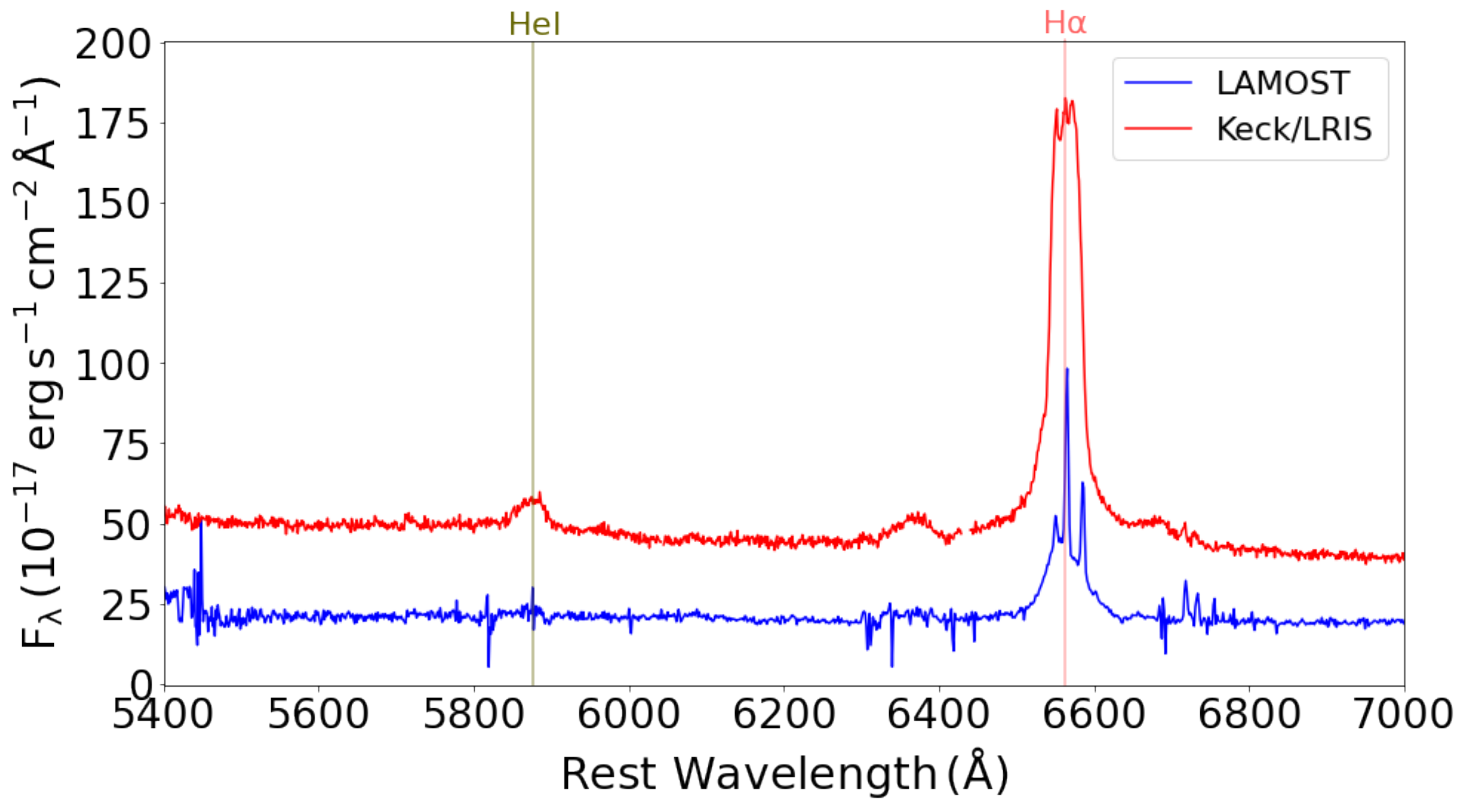}
    \caption{Comparison of the observed archival LAMOST spectrum and the raw Keck/LRIS classification spectrum. We present the regions around \hb\ (left) and \ha\ (right) lines. We see a clear appearance of a  feature around 4680 \AA\ in the Keck spectrum, which we interpret as BF, as well as the clear appearance of the \Hdelta\ line. We also find that there is enhanced emission for HeI at $\sim5875$ \AA. There are also significant signs of enhanced emission at $\sim 6375$\AA, which corresponds to the higher ionization coronal line Fe\,\textsc{x}\,$\lambda$6375.}
    \label{fig:lamostkeck}
\end{figure*}

X-ray observations with the X-Ray Telescope \citep[XRT;][]{Burrows2005} on \swift\ were obtained simultaneously with the UVOT observations. Using \texttt{XIMAGE} to process the \swift/XRT images, we find no significant X-ray detection down to $3\sigma$ at the position of AT\,2021loi. We then use \texttt{XIMAGE} to calculate the corresponding flux upper limits for each of the XRT epochs, using a circular aperture with a radius of $\sim$47\arcsec\ centered on the (optical) position of AT\,2021loi (we verify that no neighboring X-ray sources are detected within that aperture). The $3\sigma$ upper limits on the X-ray flux are between 0.006 and 0.02 counts s$^{-1}$ for the standard $0.2-10\,\kev$ XRT energy range. We use these to calculate the $2-10\,\kev$ flux upper limits, using \texttt{WebPIMMS}\footnote{\url{https://heasarc.gsfc.nasa.gov/cgi-bin/Tools/w3pimms/w3pimms.pl}}, assuming a power law photon index of $\Gamma=1.8$, which is typical of low-redshift AGNs \citep[e.g.;][]{Ricci2017}. The upper limits we derive are in the range $F(2-10\,\kev)=(1.4-4.6)\times10^{-13}\,\ergcms$. Given the source redshift, these translate to luminosity upper limits of $L(2-10\,\kev)<(2.4-7.8)\times 10^{42}\,\ergs$.

We also initiated radio measurements with with the Arcminute Microkelvin Imager-Large Array \citep[AMI-LA;]{Zwart2008,Hickish2018} and the Karl G. Jansky Very Large Array (VLA). AMI-LA is a radio interferometer comprised of eight, 12.8\,m diameter, antennas producing 28 baselines that extend from 18\,m up to 110\,m in length and operate with a 5\,GHz bandwidth, divided into eight channels, around a central frequency of 15.5\,GHz.
AT\,2021loi was first observed in the radio regime, at 15.5\,GHz, with the AMI-LA on 2021 June 15, i.e. 39 days after optical discovery, but it was not detected. AMI-LA follow-up in the same band out to 413 days after optical discovery also resulted in non-detections. The AMI-LA observations were reduced using a customized AMI-LA data reduction software package \citep{Perrott2013} and the $3\sigma$ flux density upper limits are in the range of 0.10 to 0.21 mJy, which correspond to radio luminosity limits of $\nuLnu(15.5\, \rm GHz) < (2.6-5.6) \times 10^{38}\,\ergs$.

We observed the field of AT\,2021loi with VLA on 2021 December 1, i.e. 208 days after the initial optical detection. Images were produced using the \texttt{CASA} \citep{McMullin2007} task \texttt{CLEAN} in interactive mode. While our $S$-band (3\,GHz) image show no source above the $3 \sigma$ rms limit (0.096\,mJy), the $C$- and $X$-band images (5 and 10\,GHz, respectively) show a source at the phase center, which we fit with the \texttt{CASA} task \texttt{IMFIT}. We estimate the peak flux density error to be a quadratic sum of the error produced by the \texttt{CASA} task \texttt{IMFIT} and an additional 10\,\% calibration error. 
The flux densities at 5\,GHz and 10\,GHz correspond to 0.0578$\pm$0.0071 and 0.0655$\pm$0.0087\,mJy, respectively. These, in turn, translate to radio luminosities of $\nuLnu(5\, \rm GHz) = (4.9\pm 0.6)\times 10^{37}\,\ergs$ and $\nuLnu(10\, \rm GHz) = (1.1\pm 0.2)\times 10^{37}\,\ergs$. 
Therefore, the new VLA detections correspond to a relatively weak compact radio source with a somewhat flat spectrum, and are consistent with a normal (persistent) radio-quiet AGN \citep[e.g;][]{Panessa2019}. Given the radio data in hand and previous studies (e.g.; \citealt{Smith2020} and references therein) on the correlation between radio and X-ray luminosity in AGN, we find that the radio limits and faint detections are in agreement with the upper limits for the X-ray luminosity from \swift.

AT\,2021loi is not detected in archival (NRAO\footnote{National Radio Astronomy Observatory} VLA Sky Survey; NVSS; \citealt{Condon1998}) radio survey data, which is complete only down to (roughly) 2.5\,mJy at 1.4\,GHz, and it is outside the footprint of the (slightly) more sensitive  Faint Images of the Radio Sky at Twenty-Centimeters \citep[FIRST;][]{Becker1994} survey.
Thus, we cannot say whether the new VLA detections probe an enhanced, dimmed, or persistent radio emission. 

\subsection{Optical Spectroscopy}
\label{obsspec}

We retrieved an archival spectrum of WISEA J010039.62+394230.3, the host galaxy of AT\,2021loi, from the fifth data release of the Large Sky Area Multi-Object Fiber Spectroscopic Telescope survey  \citep[LAMOST/DR5;][]{Zhao2012} obtained on 2013 December 22.\footnote{The LAMOST archive includes a yet earlier spectrum of this source, obtained two years earlier (2011 December 28). That spectrum also shows broad Balmer emission lines, similar to the LAMOST spectrum we use. However, it is significantly affected by spectral calibration issues, including residual (sky) spectral features and uncertain spectrophotometry. We thus decided to only use the latest LAMOST spectrum.} 
The LAMOST spectrum covers the range 3500 -- 9000 \AA\ with a resolution of $R=1800$. The archival spectrum shows AGN features, in particular prominent broad Balmer emission lines and strong narrow \OIII.

The classification spectrum of AT\,2021loi \citep{tns_classification} was obtained with the LRIS on Keck I on 2021 June 07, which covers the wavelength range between 3400 \AA\ and 10300 \AA\ with a resolution of $R=600-1000$. This spectrum is shown in Figure~\ref{fig:lamostkeck}, together with the archival LAMOST spectrum.

In addition, Figure~\ref{fig:lamostkeck_norm_oiii} shows the two spectra normalized in a way that would yield a common integrated \OIII\ line flux level, so to further emphasize the changes in the continuum and broad line emission (see Appendix~\ref{app:fig_lamostkeck_norm}).

We initiated a spectroscopic monitoring campaign for AT\,2021loi using the FLOYDS spectrograph \citep{Sand2011} mounted on the robotic 2m telescope at Haleakal\=a, Hawaii (which is also part of the Las Cumbres Observatory network). During the 16 months of spectroscopic monitoring, we obtained 20 spectra. FLOYDS covers the 3500--10000 \AA\ range in a single exposure by capturing two spectral orders simultaneously, yielding a spectral resolution of $R\sim400$. The exposure time of the spectra varies between 1500 and 1800 seconds and we have used a slit width of 2\arcsec. 

\begin{figure*}
    \centering
    \includegraphics[width=0.9\textwidth,height=0.9\textheight]{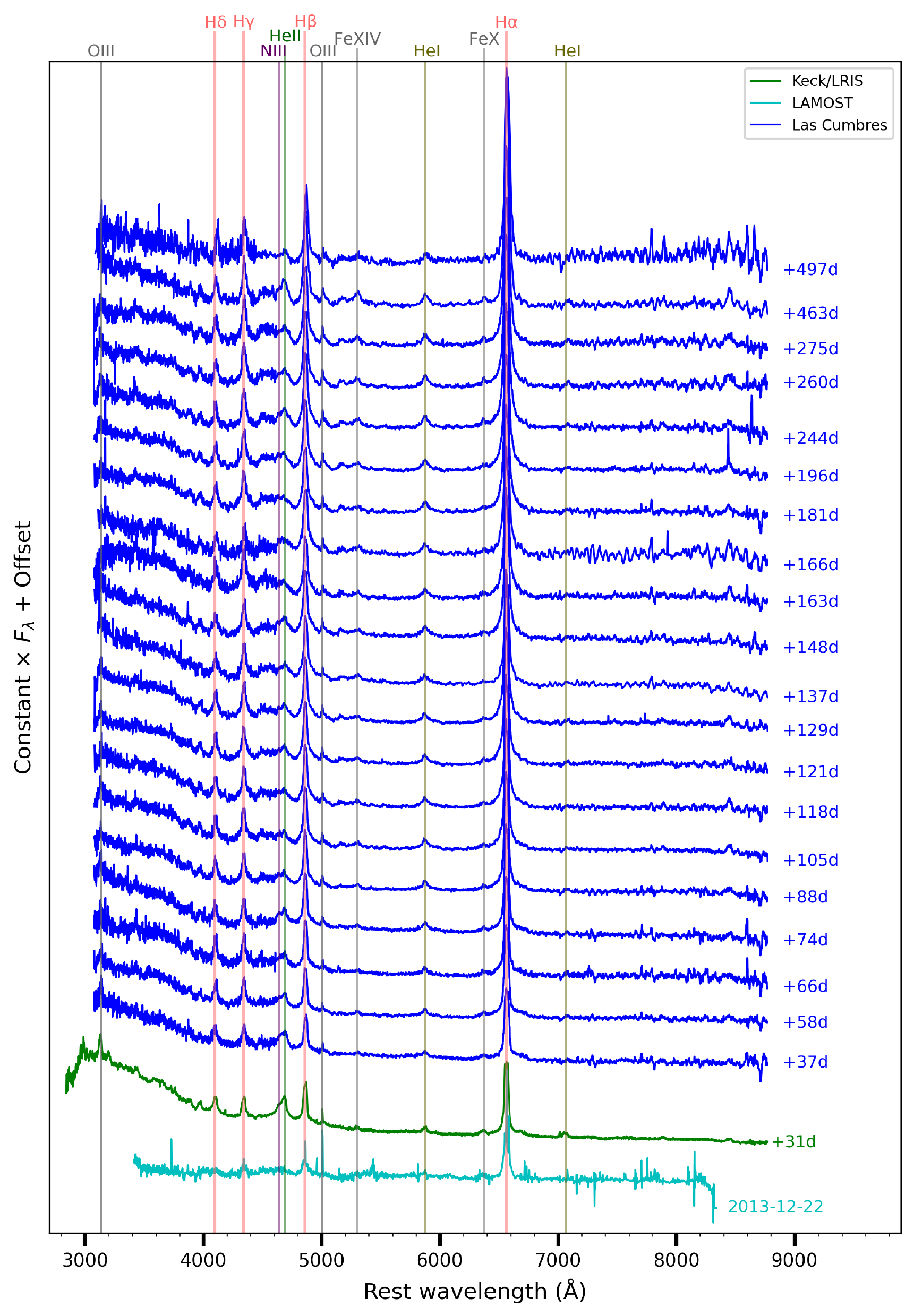}
    \caption{Spectral sequence of AT\,2021loi together with the LAMOST pre-flare spectrum. Phases are noted in observed days from discovery. Persistent, for over a year, emission features include broad Balmer lines, a feature around 4680 \AA\ as well as \hei\ emission at 5875 \AA. We also see strong, persistent \oiii\ emission at 3133 \AA\, for at least four months after the transient detection.}
    \label{fig:specseq}
\end{figure*}

The Las Cumbres Observatory spectra were reduced with a custom data reduction pipeline\footnote{\url{https://github.com/cylammarco/FLOYDS\_pipeline}} built with the \textsc{iraf}-free \textsc{python}-based \textsc{aspired} toolkit\footnote{\url{https://github.com/cylammarco/ASPIRED}} \citep{2021arXiv211102127L, lam_marco_c_2022_6903357}. The spectral images were each cropped into a red and a blue image. Standard data reduction procedures were applied to trace and then optimally extract the spectral information using the \citet{1986PASP...98..609H} algorithm. Wavelength calibration was performed using the built-in calibrator powered by \textsc{rascal}\footnote{\url{https://github.com/jveitchmichaelis/rascal}} \citep{2020ASPC..527..627V, veitch_michaelis_joshua_2021_4124170}. Standard stars from the same night were used for flux calibration when available; otherwise, the ones observed closest in time to the science observations were used. Finally, atmospheric extinction and telluric absorption were removed. We present all spectra in Figure \ref{fig:specseq}.

Basic details regarding all the optical spectra of AT\,2021loi used in this work are provided in Table~\ref{tab:opt_spec} (in the Appendix), along with key spectral measurements (see Section~\ref{subsec:specev}).

\section{Analysis}
\label{analysis}

\subsection{Photometric Evolution}
\label{subsec:photev}


The optical brightness of AT\,2021loi increased by a factor of $\approx4$ between 2021 May 7 (first detection by ZTF) and 2021 June 22, from the initial $M_r = -19.04\pm0.15$ to peak magnitudes of $M_r = -21.08\pm0.03$ and $M_g = -20.98\pm0.04$ (all reference-subtracted). The latter corresponds to $\nuLnu(g) \simeq 7.8\times 10^{43}\,\ergs$.  Following the peak, the optical lightcurve shows a slow, steady decline until mid-June 2022 (Figure \ref{fig:lc}). 
We fit the decline phase between the main peak and the re-brightening phase ($59386 <{\rm MJD} < 59750$) with a power-law of the form $F \propto (t-t_0)^\alpha$. We find a power-law slope of $\alpha \simeq -0.48$ if we fix the value of $t_0$ to the discovery time of the optical flare, and $\alpha \simeq -0.78$ if we let $t_0$ (which yields $t_0=59253\pm40$) be a free parameter, with formal uncertainties of ${\approx}0.01$ and $0.1$, respectively. 
In any case, these power-laws are much shallower than the $\alpha=-5/3$ associated with TDEs \citep[e.g.;][]{Rees1988}, which does not provide a good fit (especially for the $r$ band) for the data (Figure \ref{fig:lc}) . 
After mid-June 2022, AT\,2021loi experiences a rebrightening and shows a second peak in the ZTF $r$-band at $M_r=-20.59\pm0.05$ mag on 2022 July 15, about 390 days after the main peak. This second peak is also seen in the ZTF $g$ and ATLAS $o$ and $c$ bands. An observational gap did not allow Las Cumbres $gri$ bands to cover the rebrightening. The decline after the second peak, although faster ($\alpha\simeq-0.9$, $r$ band), it is still very different to that of TDE.

We compare the $uvm2$ UVOT observations to the archival GALEX (NUV; $\lambda_\textrm{eff} = 2304$\,\AA) data as $uvm2$ ($\lambda_\textrm{eff} = 2245$\,\AA) is the most relevant UVOT filter for such a comparison. In the first UVOT observation, which also resulted in the highest NUV flux we obtained during our monitoring campaign, we measure a $uvm2$ absolute magnitude of $M_{uvm2}=-21.06\pm0.07$ which corresponds to $\nu L\nu (NUV) = 2.3\times 10^{44}\,\ergs$. This is a factor of $\sim$20 increase from the archival GALEX detection, where the $M_{\rm NUV} = -17.84\pm0.14$ magnitude translates to a $\nuLnu(UV) = 1.1\times 10^{43}$\,\ergs. Following this measurement, the NUV flux shows a gradual decline before the rebrightening mentioned above. The monochromatic NUV luminosity we measure $\sim$14 months after the initial transient detection (and one year after the initial \swift\, detection) is $\nuLnu\simeq5.7\times 10^{43}\,\ergs$, i.e., a drop of $\sim$ 75\% but still higher than the archival GALEX measurement by a factor of $\sim$5. 

As already mentioned in Section~\ref{sec:phot_data}, the MIR lightcurves (Figure \ref{fig:ltlc}, top panel) show limited variability prior to the optical flare, and a noticable flux increase in the first MIR measurement that follows the optical flare. This enhanced MIR emission is separated by at most ${\sim}64$ days from the detection of the optical outburst or ${\sim}30$ days compared with the peak optical emission (both in the rest-frame).
These timescales are mere upper limits on the real delay, driven by the (low) cadence of the NEOWISE-R monitoring.
To put these timescales in context, we note that the inner edge of the dusty torus, i.e. where the temperature reaches the sublimation threshold, is expected to be ${\sim}100-300$ light days away from the central engine, based on our near-peak \Lbol\ estimate for AT\,2021loi. 
This range is driven by the range in possible dust grains (see Equations 2 \& 3 in \citealt{Mor2009}) and is supported by NIR reverberation mapping campaigns \cite[e.g.][]{Suganuma2006,Koshida2014}.

We can thus infer that, given the data in hand, the increase in MIR emission is broadly consistent with the reprocessing of the (sharp) rise in UV/optical radiation by circumnuclear dust, in a way that is not markedly different from what is seen in normal (persistent) AGNs.
From the NEOWISE-R lightcurves, there is also some variability in the $W1-W2$ color, which changes from around 0.66 mag prior to the outburst, to $\sim$0.40 mag during the transient state. 
While the former, pre-outburst MIR color is close to the cut used to identify AGN (i.e., $\gtrsim0.7$), the latter post-flare color is in fact, less consistent with what is found for persistent AGNs.
This may not be surprising, given the possibility that the emission from either the exceptionally UV-bright flare and/or the hottest, pure-graphite part of the torus is enhancing the shorter-wavelength $W1$ band (i.e. relative to normal AGNs).

There is no ROSAT counterpart to WISEA J010039.62+394230.3 in the all-sky survey conducted during 1990--1991. From this we infer an archival $3\sigma$ X-ray upper-limit of $F(0.1-2.4\,\kev)\lesssim 10^{-13}\,\ergcms$ \citep[][]{Boller2016}, which translates to $L(0.1-2.4\,\rm keV) \lesssim 2\times 10^{42}\,\ergs$ but also to $L(2-10\,\rm keV) \lesssim 2.5\times 10^{42}\,\ergs$. 
In order to compare this archival ROSAT upper limit with the post-flare XRT upper limits described in Section~\ref{subsec:xrayradio}, we utilize the observed relation between X-ray and NUV emission in persistent AGNs, which is commonly quantified through the anti-correlation between the optical-to-X-ray spectral slope, $\aox \equiv \log(f_\nu [2\,\kev] / f_\nu [2500\,\textrm{\AA}]) / \log(\nu[2\,\kev]/ \nu [2500\,\textrm{\AA}])$, and the monochromatic NUV luminosity \cite[see, e.g.,][and references therein]{Just2007,Lusso2016,Nanni2017}.

Specifically, we rely on the relation derived in Eq.~2 of \cite{Nanni2017}, of the form $\aox = -0.155 \log(L_{2500\,\textrm{\AA}})+3.206$.
From the (extinction corrected) archival GALEX NUV magnitude of AT\,2021loi, $m_{\rm NUV}=19.65$, we derive an archival NUV luminosity of $L_\nu = 8.5\times10^{27}\,\ergs\,\textrm{Hz}^{-1}$. Plugging this into the \cite{Nanni2017} relation, the expected spectral slope is $\aox = -1.12$, which in turn yields an intrinsic monochromatic X-ray luminosity at 2\,\kev\ of $\nuLnu(2\,\kev) = 5.6\times 10^{42}\,\ergs$. Extrapolating the 2\,\kev\ fluxes to the 2--10\,\kev\ range assuming a photon index of $\Gamma = 1.8$, we derive an expected X-ray luminosity of $L(2-10\,\kev) \simeq 6\times 10^{42}\,\ergs$, which is a factor of $\sim$3 higher than the upper limit from ROSAT. 
We stress that this estimate for the expected X-ray emission of AT\,2021loi carries significant uncertainties. In particular, a $1\sigma$ uncertainty of ${\approx}0.2$ on $\aox$ at any given $L_{2500\,\textrm{\AA}}$ (consistent with what is reported in \citealt{Nanni2017}) would translate to an uncertainty of almost 0.6 dex on the expected X-ray luminosity we derive through this approach (i.e., a factor of $\sim$4).\footnote{Other sources of uncertainty include the possible range of $\Gamma$ \citep{Ricci2017} and the fact that some AGNs are known to be ``X-ray weak'' (i.e. have significantly lower \aox\ given their NUV emission; see \citealt{Luo2015} and references therein).}
Following the same approach for the near-peak \swift\ NUV measurement, of $m_{\rm NUV}=16.82$, yields $\aox \simeq -1.33$ and $L(2\,\kev)=3.1\times 10^{43}\,\ergs$. 
Extrapolating the 2\,\kev\ flux to the 2--10\,\kev\ range using $\Gamma = 1.8$, we derive an expected X-ray luminosity of $L(2-10\,\kev) = 5.0 \times 10^{43}\,\ergs$. The upper limits from \swift/XRT correspond to $L(2-10\,\kev) < (2.4-7.8)\times 10^{42}\,\ergs$. 
This means that if the broad-band SED of AT\,2021loi would have followed that of persistent broad-line (unobscured) AGNs, then---based on the near-peak NUV emission---the source should have been detected in the X-rays, while in practice, it was not.
We discuss this further in Section \ref{subsec:diss_bffs}.

\begin{figure*}
    \centering
    \includegraphics[width=1.0\textwidth]{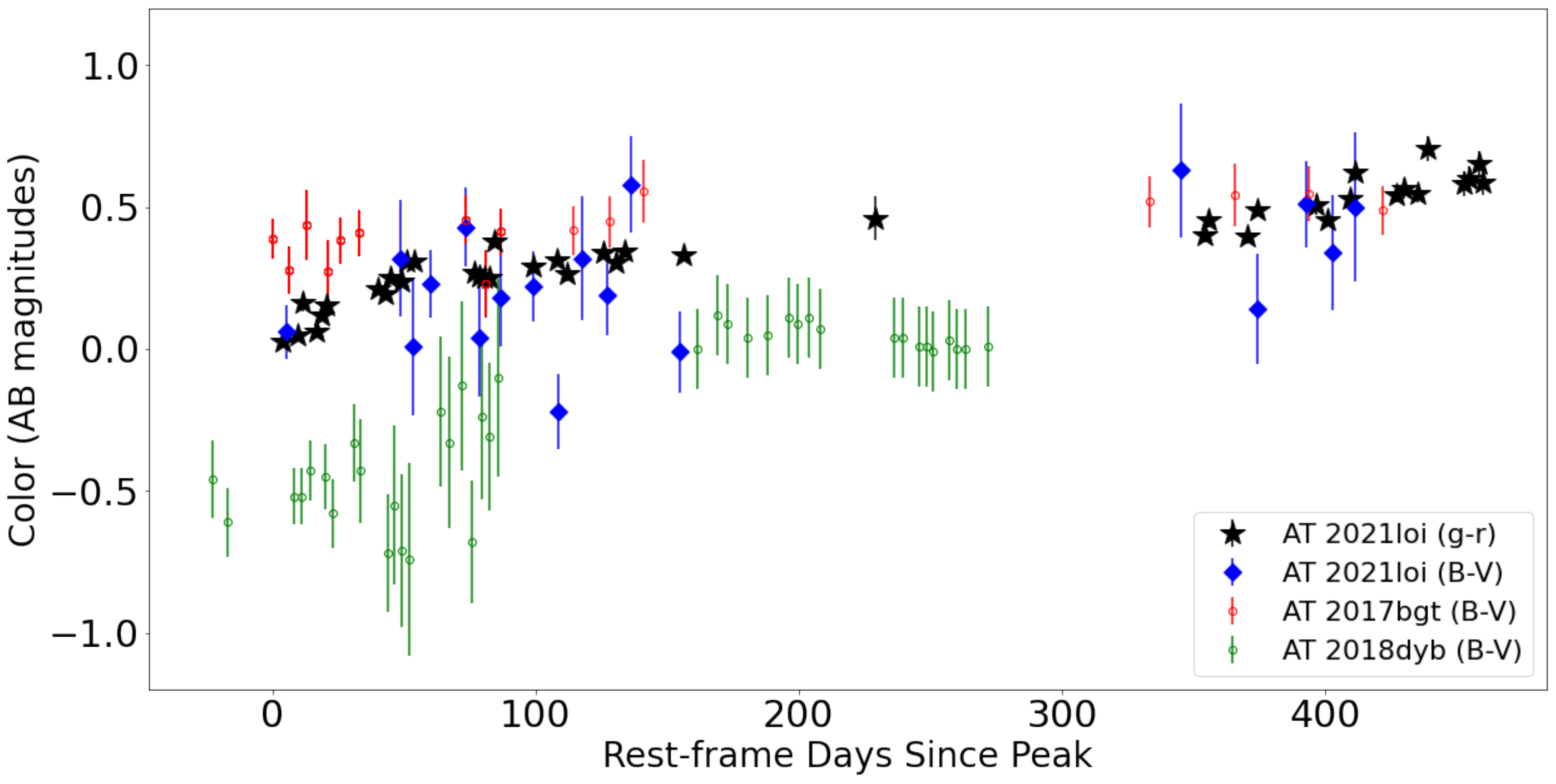}
    \caption{Color evolution ($g-r$ and $B-V$) of AT\,2021loi, the BFF AT\,2017bgt, and the BF TDE AT\,2018dyb. AT\,2021loi is redder around the peak than AT\,2018dyb. AT\,2021loi also shows very limited color evolution, similar to AT\,2017bgt.}
    \label{fig:colourev}
\end{figure*}

Figure \ref{fig:colourev} shows the color evolution of the UV bright AT\,2021loi compared to that of the BFF AT\,2017bgt \citep{Trakhtenbrot2019a} and the BF TDE AT\,2018dyb \citep{Leloudas2019}. The color of AT\,2021loi shows limited evolution similar to that of AT\,2017bgt. Also, AT\,2021loi is redder compared to AT\,2018dyb.

\begin{figure}
    \centering
    \includegraphics[width=\linewidth]{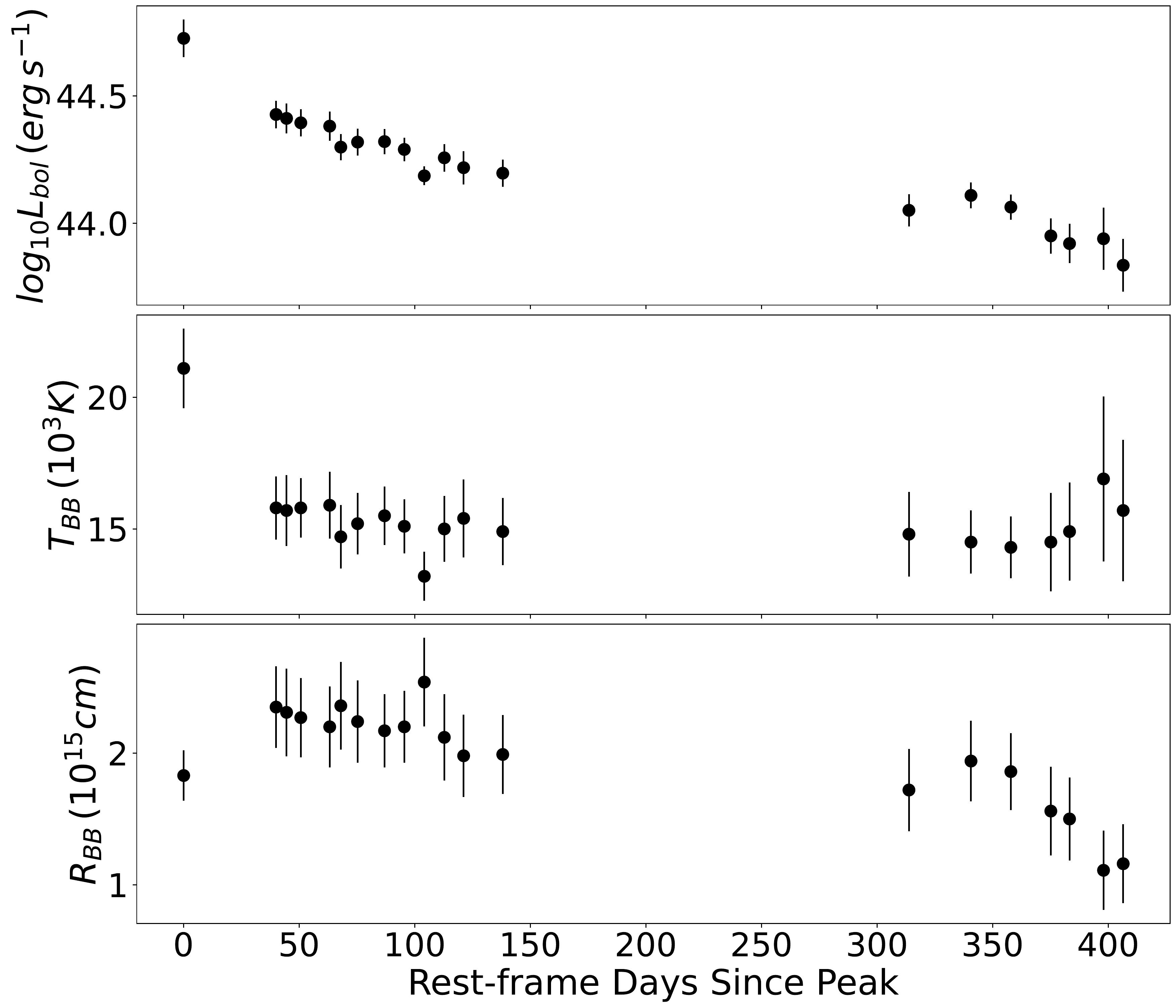}
    \caption{The bolometric luminosity (top), temperature (middle), and radius (bottom) from fitting a blackbody to the UV and optical photometry.}
    \label{fig:bb}
\end{figure}

We fit a blackbody to the ZTF ($g$ and $r$), ATLAS ($c$ and $o$), LCO ($g$ and $i$) and all the \swift\ photometry for each epoch in which we have \swift\ data (linearly interpolating neighboring optical epochs), using the blackbody fitting tool \textsc{superbol}\footnote{\url{https://github.com/mnicholl/superbol}}\citep[][]{Nicholl2018}. We then calculate the bolometric luminosity by integrating each observed SED, approximating the missing flux outside the observed bands, using the blackbody fits. 
Our results are presented in Figure \ref{fig:bb}.
The peak blackbody temperature (i.e., at UV peak but also the peak temperature) is $(21.1\pm 1.5)\times 10^3\,\textrm{K}$. After $\sim43$ days (in rest-frame) it declines to $(15.8\pm1.2)\times10^3\,\textrm{K}$, eventually reaching $(14.4\pm1.8)\times10^3\,\textrm{K}$. At these temperatures, we are sampling the blackbody continuum fully, especially with the UV data \citep{Arcavi2022}. %

Compared to a large compilation of TDEs \citep{VV2020}, AT\,2021loi is on the low-temperature end. Moreover, AT\,2021loi has a steady slow decline, unlike several TDEs, which show a decline followed by a rise in temperature. In terms of inferred radius, AT\,2021loi has larger blackbody radii (1.8--2.6$\times10^{15}\,\textrm{cm}$) compared to the vast majority of the \cite{VV2020} TDEs (typical $\lesssim6\times10^{14}\,\textrm{cm}$) and shows very little evolution. In contrast, TDE radii decline a few weeks after peak. The peak bolometric (blackbody) luminosity of AT\,2021loi is $L^{\rm{BB}}_{\rm{bol} }  = 4.8 \times 10^{44}\,\ergs$, which is higher than the majority of the TDEs presented in \cite{VV2020}. In comparison, the estimated temperature at the peak of the OGLE17aaj BFF is $4\times10^4\, K$, which is much higher than that of AT\,2021loi. \citep{Gromadzki2019}. 

\subsection{Spectroscopic Evolution}
\label{subsec:specev}

\begin{figure*}
    \centering
    \includegraphics[width=0.95\textwidth]{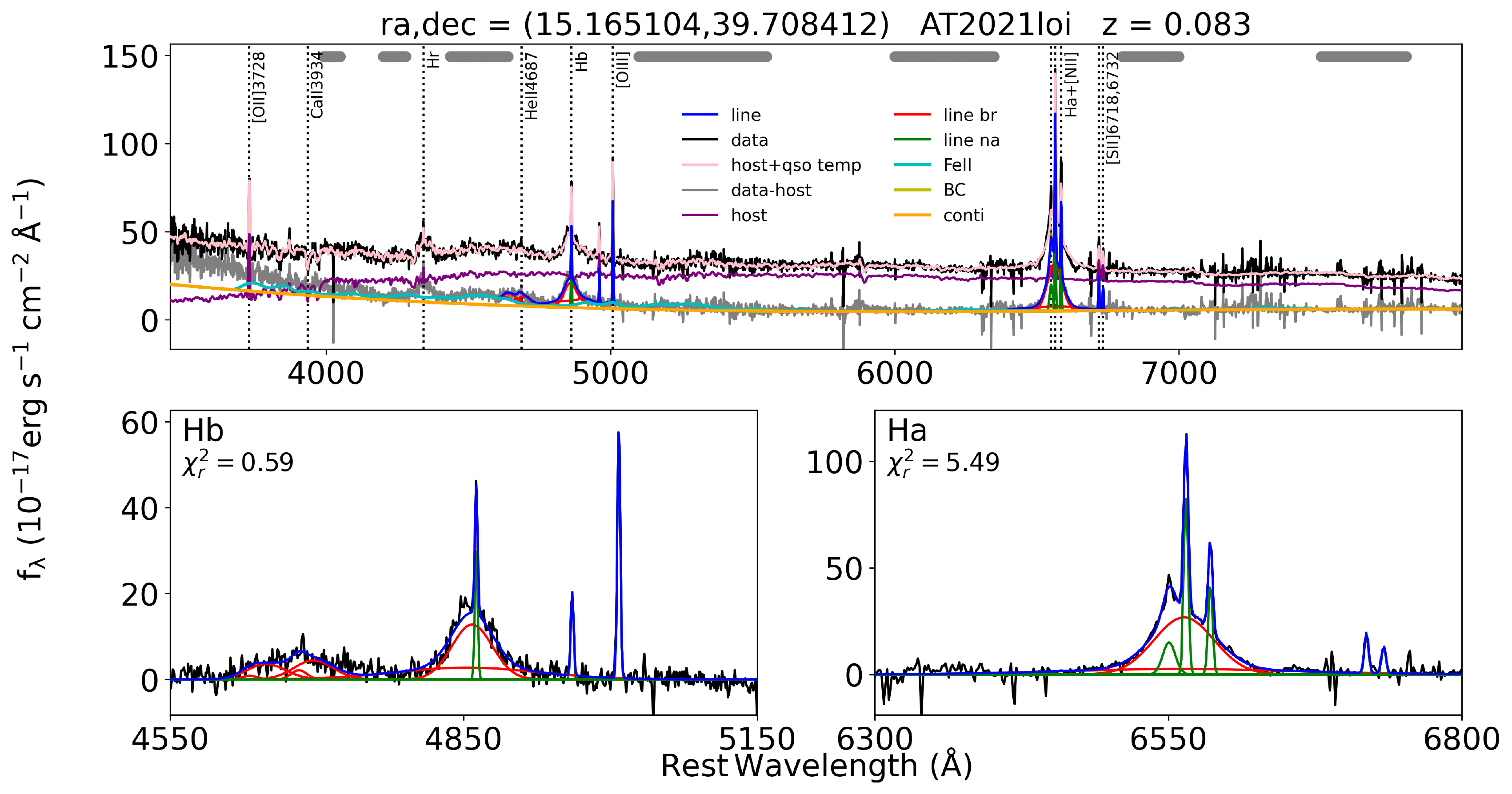}
    \caption{Decomposition of the archival LAMOST spectrum with \textsc{pyqsofit} along with the best-fit model and the model components. Gray horizontal lines on the top show the windows used for the continuum estimate. The black line represents the initial data, whereas the gray line represents the host subtracted data. The host is plotted as the purple line. The best-fit emission lines, as a sum of narrow (green line) and broad lines (red line), are marked in blue. The bottom subplots are a zoomed version of the components for \ha\, and \hb.}
    \label{fig:lamostfit}
\end{figure*}

We use the \textsc{pyqsofit} fitting procedure\footnote{\url{https://github.com/legolason/PyQSOFit}} \citep{Guo2018} to model both the archival LAMOST spectrum, the first flare Keck spectrum and the FLOYDS spectra. \textsc{pyqsofit} fits and decomposes the continuum emission (with AGN and host galaxy components) and fits the broad and narrow emission lines. The continuum is fitted as a simple power law, and we add a polynomial component when fitting fails without it. 
All narrow emission lines are modeled with a single Gaussian, whereas the broad lines are modeled with either one or two Gaussians.
We initially tried to model all broad emission lines with a single (broad) Gaussian, which works for simpler and/or weaker broad line profiles (e.g. \Hdelta). However, we found that some line profiles (e.g., \Hbeta) are more complex and two broad components are required to achieve a satisfactory fit. These multiple broad Gaussians are not constrained in terms of offset velocity, peak intensity, or any other parameter.
We stress that the choice to use either one or two broad Gaussian profiles was applied consistently to all spectra (i.e., all epochs).
This fitting approach is in line with common practice in fitting the (sometimes complex) broad line emission profiles in normal, persistent AGNs \cite[see, e.g.,][]{GreenHo2005,Shen2011,TrakhtenbrotNetzer2012,MejiaRestrepo2018,Rakshit2020}. 
We fit the 4500 to 5100 \AA\ region, which covers the entire \NIII\ and \HeIIop\ region as well as \hb\ and the [\oiii]\,$\lambda\lambda$4959,5007 lines. In addition, we fit the \ha\, region, including the $[\textrm{N\,\textsc{ii}}]\,\lambda\lambda$6548,6584 and  $[\textrm{S\,\textsc {ii}}]\,\lambda\lambda$6718,6732 doublets. 

We first fit the archival spectrum from LAMOST with all the available components within \textsc{pyqsofit} (i.e., the host component, the power law continuum, the Balmer continuum, and the iron emission). The fit results are presented in Figure \ref{fig:lamostfit}. This spectrum shows broad Balmer emission lines with $\fwha$ and $\fwhb \simeq 3000\,\kms$, thus clearly capturing a broad-line, unobscured AGN. In addition, it also shows strong narrow \OIII\ emission. The key diagnostic narrow line ratios are $\log([\oiii]\,\lambda\,5007/\hb) \simeq 0.31$ and $\log(\nii\,\lambda\,6583/\ha)\simeq -0.30$, which would correspond to the composite region in the BPT \citep{BPT1981} - i.e. both star formation and AGN activity are responsible for producing the narrow lines in this galaxy \citep{Kewley2006}. 
Table~\ref{tab:opt_spec} (in the Appendix) lists some key spectral measurements resulting from our analysis.

Based on relations between luminosity and broad line region (BLR) size derived from reverberation mapping campaigns, the LAMOST spectrum yields a BLR size of $\RBLR \simeq 2.9$ and 5.4 light days using the prescriptions of \cite{TN2012} and \cite{Bentz2013} respectively. We use the $L(\ha)$ luminosity and the respective \fwha\ to calculate the SMBH mass following Equation 2 in \cite{MR2022} and find $\mbh=1.0\times10^7 \Msun$. 
The unncertainties of such mass estimates are of the order 0.3-0.5 dex, including  uncertaninties on both the $\RBLR-\Lop$ relation and the so-called virial factor \cite[see, e.g.,][and references therein]{Shen2013,MR2022}.
We estimate the bolometric luminosity (\Lbol) from the host galaxy subtracted monochromatic luminosity at rest-frame 5100\,\AA\ ($3.2 \times10^{42}\,\ergs$) and a bolometric correction of $f_{\rm bol} (5100\,\AA) \equiv \Lbol/\Lop = 9.26$ \cite[e.g.,][]{Shen2008, MacLeod2010}. This estimate has an uncertainty of order $\sim0.3$ dex, due to the uncertainty on the bolometric correction  \cite[e.g.,][and references therein]{Runnoe2012,Duras2020}. 
From \Lbol\ and \mbh, we derive an estimate for the pre-flare Eddington ratio of $L/L_{\rm Edd}\approx0.02$ (with large systematic uncertainties, stemming from both \Lbol\ and \mbh).
Despite the large uncertainties associated with this Eddington ratio estimate, we conclude that the source, in its pre-flare state, was well within the sub-Eddington accretion regime. 

\begin{figure*}
    \centering
    \includegraphics[width=0.95\textwidth]{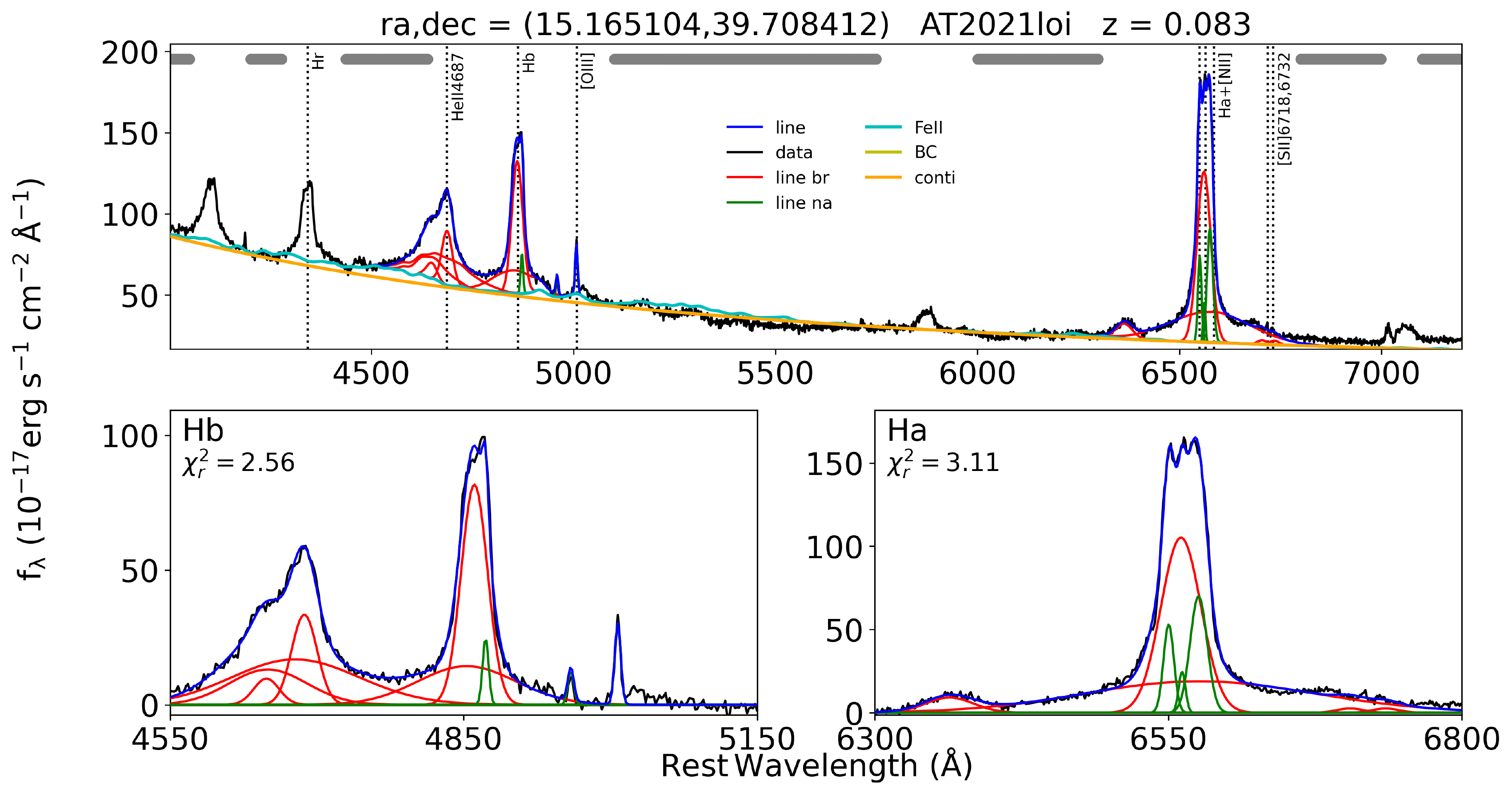}
    \caption{Decomposition of the Keck spectrum with \textsc{pyqsofit} along with the best-fit model and the model components. Grey horizontal lines on the top show the windows used for the continuum estimate. The black line represents the host subtracted data. The best-fit emission lines, as a sum of narrow (green line) and broad lines (red line), are marked in blue. The bottom subplots are a zoomed version of the components for \ha\, and \hb.}
    \label{fig:keckfit}
\end{figure*}

The earliest spectrum after the flare, which is also the one nearest the optical peak time, is the Keck/LRIS spectrum obtained 31 days after discovery. Since the Keck spectrum has a lower resolution compared to LAMOST, we use the host component as measured for the LAMOST spectrum by \textsc{pyqsofit} to apply host subtraction to the Keck spectrum. After that, we process the Keck spectrum with \textsc{pyqsofit} (Figure \ref{fig:keckfit}) without applying host decomposition but only continuum fitting, iron fitting, and emission lines fitting.

For the Keck spectrum, we find $ \Lop =3.0 \times10^{43}\,\ergs$. From this, along with the \mbh\ estimated from the LAMOST spectrum and the same bolometric correction as mentioned above, we derive an Eddington ratio at the epoch of the Keck spectrum (i.e. near-peak) of $\lledd\approx0.2$. This is an increase by a factor of $\sim$10 compared to the pre-flare phase, but it is still well within the sub-Eddington regime. 
The UV peak luminosity found in section \ref{subsec:photev}, combined with a UV bolometric correction of $f_{\rm bol} ({\rm UV})=3.5$ \cite[see][]{Kaspi2000, Netzer2016}, yields a bolometric luminosity of $7.7 \times 10^{44}\,\ergs$, which in turn gives $\lledd\simeq 0.5$. The higher value of UV-based \Lbol\ estimate is a result of the UV-bright nature of the transient. Although more than twice higher than what is found from \Lop, the UV-based \lledd\ estimate is still within the sub-Eddington regime, although we note that the large systematic uncertainty on this estimate means it is formally consistent with $\lledd \approx 1$. 
The bolometric luminosity generated from the blackbody fit in the previous section is $4.8 \times 10^{44}\,\ergs$, which is lower than the one estimated from the UV flux, resulting in an even lower Eddington ratio.
In any case, notwistanding the range of \lledd\ estimates and the significant systematic uncertainties associated with them, we can still conclude that the data in hand does not support a scenario in which the AT\,2021loi flare approached the Eddington limit of the accreting SMBH. 

\begin{figure*}
    \centering
    \includegraphics[width=0.45\textwidth]{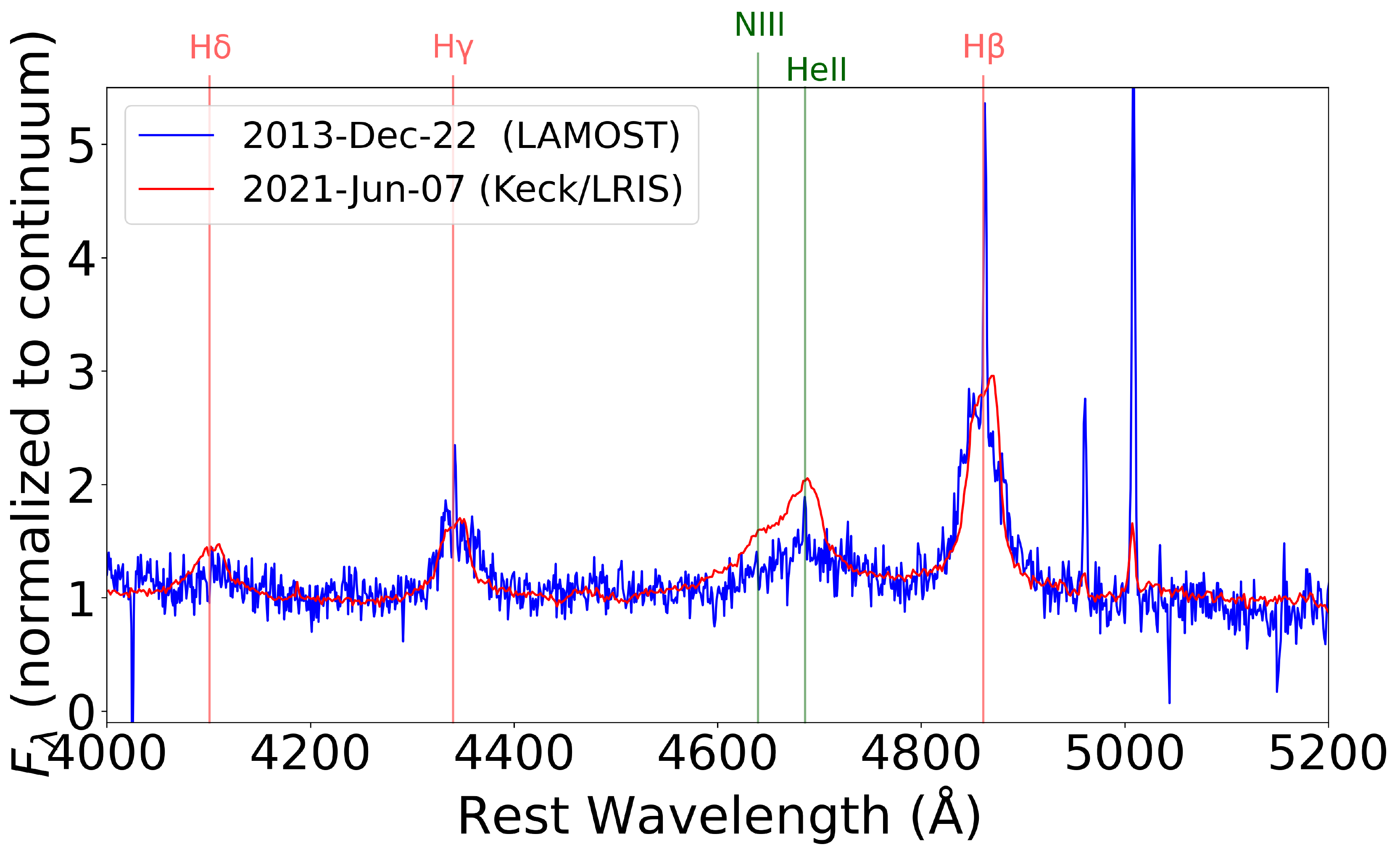}
    \includegraphics[width=0.45\textwidth]{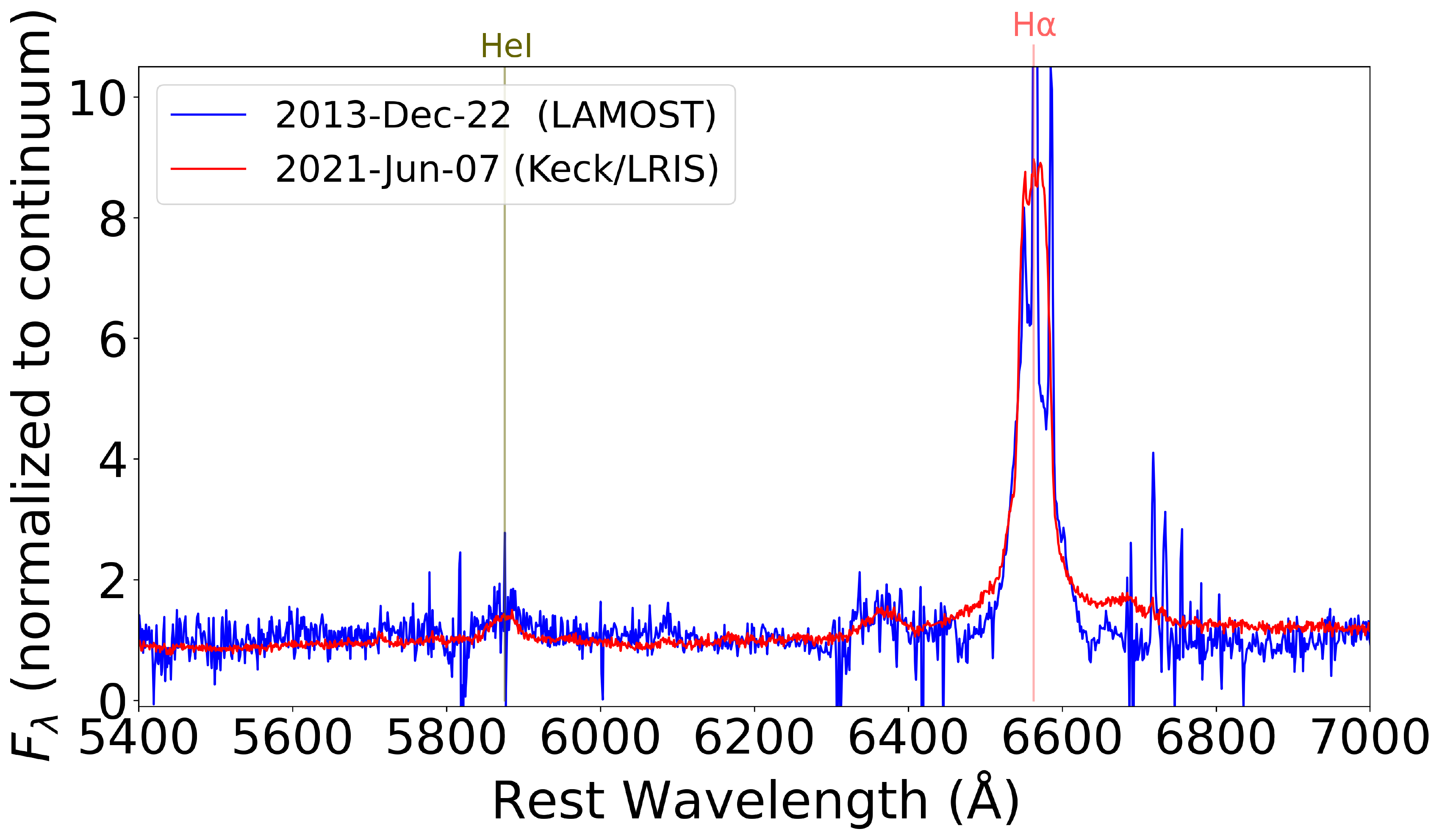}
    \caption{Comparison of the host-subtracted and extinction corrected LAMOST and Keck spectra divided by the power-law continuum and the iron complex as fitted by \textsc{pyqsofit}. We see that there is indeed enhanced emission in the region around 4680 \AA\ but also for \Hdelta, with regards to   \Hbeta. The \halpha\ remains essentially unchanged.}
    \label{fig:lamostkeck_norm}
\end{figure*}

In Figure \ref{fig:lamostkeck_norm}, we plot the LAMOST and Keck spectra divided by the power-law continuum and the iron component as fitted by \textsc{pyqsofit} in order to see the difference in the strength of the emission lines with respect to the continuum and iron complex. We see that the relative strength of the feature around 4680\AA\ to \hb\ has increased.
In order to quantify the changes in the \HeIIop+\NIII\ region between the LAMOST pre-flare and Keck (near optical peak) spectra, we perform the following procedure on both spectra. We start by fitting both the \HeIIop\ and \NIII\ complex with two Gaussian components for each transition, forcing the widths to match that of the \hb\ broad component. Each of these broad components is allowed to have a FWHM in the range of $2000-10000$ \kms. In the case of LAMOST, although it is possible to fit the emission feature around 4680 \AA\, ($\textrm{FWHM} > 6000\,\kms$) as either one or two components (i.e. only \HeIIop\, or \HeIIop\ blended with \NIII), it is not clear if the feature indeed consists of two peaks. In the case of the Keck spectrum, we see two peaks more clearly, similar to previously identified BFFs. We list the flux (measured with \textsc{pyqsofit}) ratios between emission lines of interest in Table \ref{tab:ratios}. 

We look for differences in the integrated flux density of the total \HeIIop\, and \NIII\ region relative to the integrated flux density of \hb. We do this by integrating the flux density after host, continuum, and iron subtraction between 4580 \AA\, and 4750 \AA\, (the region that covers \HeIIop\, and \NIII) and between 4800 \AA\, and 4900 \AA\, (the region that covers \hb). We use two approaches to quantify this ratio. In the first case, we integrate the observed flux in the region with no emission line fitting. We find that the ratio between the integrated fluxes in the two regions is 0.53 for the LAMOST spectrum and 1.12 for the Keck spectrum, meaning that the ratio increased by a factor of $\approx2$. Using the flux as extracted from the Gaussians fit to the emission lines, we find a similar line ratio increase from $F(\HeIIop+\NIII)/F(\hb)\sim0.58$ in the LAMOST spectrum to $F(\HeIIop+\NIII)/F(\hb)\sim1.09$ in the Keck spectrum. 

We also find that the ratio between the integrated flux without fitting the \Hdelta\ spectral region (4050--4150\,\AA) to that of \hb\ increased from $0.22\pm0.01$ in the LAMOST spectrum to $0.35\pm0.02$ in the Keck spectrum. The flux ratio of $F(\Hdelta)/F(\hb)$ from Gaussian fitting increased from $0.24\pm0.02$ in the LAMOST spectrum to $0.36\pm0.03$ in the Keck one. On the other hand, the flux ratio of \Hgamma\ to \hb\ remained consistent with $F(\Hgamma)/F(\hb)=0.51\pm0.02$ for LAMOST and $0.48\pm0.03$ for Keck (without fitting) and at $F(\Hgamma)/F(\hb)=0.48\pm0.05$ for LAMOST and $0.41\pm0.03$ for Keck (with fitting). We find, though, that the $F(\Hgamma)/F(\hb)$ ratio  remained constant, whereas the $F(\Hdelta)/F(\hb)$ significantly increased by $\sim1.5$, providing more support for the BF mechanism as the \Hdelta\ feature is likely blended with N\,{\sc iii} emission at $\lambda\lambda$4097, 4104 \citep{Netzer1985}. 

Although the spectrum appears to have a strong iron bump around 4500--4600\,\AA\ after the \textsc{pyqsofit} fitting, we find that He\,\textsc{II} at 4687\,\AA\ was present in the archival LAMOST spectrum. Typically, in AGN, $F(\HeIIop)/F(\hb)\leq0.05$ \citep{VandenBerk2001}. For the LAMOST spectrum, the intensity ratio $F(\HeIIop+\NIII)/F(\hb)$ is $\sim$0.5, which is already 10 times higher than typical AGN. For the Keck spectrum where \NIII\ and \HeIIop\ lines can be clearly separated, we find that $F(\HeIIop)/F(\hb)=0.71\pm0.09$, $F(\NIII)/F(\hb)=0.47\pm0.07$ and $F(\NIII)/F(\HeIIop)=0.66\pm0.11$. Given the $F(\HeIIop)/F(\hb)$ and $F(\NIII)/F(\hb)$ ratios seen in AT\,2021loi and the predicted relative intensities for AGN as suggested by \cite{Netzer1985}, the gas producing the BF must have a very high density ($n_H>10^{9.5}\,\textrm{cm}^{-3}$) and higher N and O abundances relative to the cosmic ones. We find that the [\hei]$\lambda$5875 flux relative to \OIII\ increased from 1.17 in the LAMOST spectrum to 2.53 in the Keck spectrum; however, the flux ratio to \hb\ has decreased from 0.27 in the LAMOST to 0.13 in Keck. 

To trace the evolution of the key spectral features of AT\,2021loi, we apply the \textsc{pyqsofit} fitting also to the lower resolution Las Cumbres/FLOYDS spectra, using the same approach as we did for the Keck spectrum (i.e., subtracting the host component as extracted from the archival LAMOST spectrum). We use the same two approaches as before, i.e., first integrating over the observed flux in the region with no emission line fitting and second, the integrated fluxes are found after Gaussian fitting for \HeIIop\ and \NIII. At the top panel of Figure \ref{fig:lc}, we plot the ratio of the total \HeIIop\, and \NIII\, integrated flux to the respective integrated flux of the broad \hb\, as a function of rest-frame time with respect to $\textrm{MJD}=59381$.  The flux is host, continuum, and iron subtracted in both cases (i.e., with and without fitting). 

The top panel of Figure \ref{fig:lc} shows that there is enhanced emission in the region around 4680 \AA\, (relative to \hb) consistent with BF brightening. This enhanced emission lasts for at least 27 rest-frame days after the Keck spectrum and 58 rest-frame days after the initial transient detection. The ratio of the flux of the BF features to \hb\, appears to become weaker than the respective ratio in the LAMOST spectrum after 35 rest-frame days from the Keck spectrum, but it is evident that the emission around 4680\AA\, still exists. The lower resolution of the FLOYDS spectra compared to LAMOST and Keck may under-predict our measured ratios due to blending with Fe features. 

We now turn to investigate the high ionization iron lines. These lines have been seen in the past in AGN and for Fe\,\textsc{x}\,$\lambda6375$ it has been found that the maximum $F([\textrm{Fe\,\sc{x}}]\lambda6375)/F(\OIII)$ ratio is $\sim0.24$ \citep{Nagao2000}. The LAMOST spectrum of AT\,2021loi shows a $F([\textrm{ Fe\,\sc{x}}]\lambda6375/F(\OIII)$ ratio of $\sim$0.6. In the Keck spectrum, we find this value to be 2.06. 


In the LAMOST spectrum, the FWHM of Fe\,\textsc{x}\,$\lambda6375$ is $\sim$1900\,\kms\, is lower than that of \hb, \ha, \hei\, ($\sim$3000\,\kms{}) and of that of the BF lines (\HeIIop+\NIII $\sim$7000\,\kms{}). For the  Keck spectrum, the Fe\,\textsc{x}\,$\lambda6375$ emission line's FWHM is $\sim$2000\,\kms. We note that this is surprising for two reasons: 1) The line seems almost unchanged between the two spectra, and 2) in both cases, the line is consistent with a BLR, whereas in \cite{Wang2012}, coronal emitters are found to have Fe\,\textsc{x}\,$\lambda6375$ with FWHM between 200 and 1000\,\kms, i.e., consistent with the narrow line region (NLR). 

\begin{deluxetable}{ccc}
\tabletypesize{\footnotesize}
\tablecolumns{2} 
\tablecaption{\label{tab:sims}Emission line ratios from the LAMOST and Keck spectra.} 
\tablehead{\colhead{Flux Ratio} & \colhead{\hspace{.13cm}Pre-flare}\hspace{.13cm} & \colhead{\hspace{.13cm} Transient state} } 
\startdata
\hline
    $F(\ha^b)/F(\hb^b)$ &  $2.33\pm0.21$& $2.12\pm0.34$ \\
    $F(\HeIIop)/F(\hb^b)$   & - & $0.71\pm0.09$ \\
    $F(\NIII)/F(\hb^b)$ &- & $0.47\pm0.07$ \\
    $F(\HeIIop+\NIII)/F(\hb^b)$ & $0.58\pm0.13$ & $1.12\pm0.05$ \\
    $F(\Hgamma)/F(\hb^b)$& $0.48\pm0.03$ & $0.41\pm0.02$ \\
    $F(\Hdelta)/F(\hb^b)$& $0.24\pm0.02$ & $0.36\pm0.03$ \\
\enddata 
\tablecomments{For the ratios on this table, we use fluxes from Gaussian fitting. The superscript `b' denotes the broad component.}
\label{tab:ratios}
\end{deluxetable}

\section{Discussion}
\label{discuss}

AT\,2021loi is an optically-discovered, UV-bright flare observed at the center of an active galaxy.
The steep rise in the optical by a afactor of $\sim4$ in about 40 days, as well as in the UV (a factor of $\sim20$ brightening),  distinguishes AT\,2021loi from the usual variability of unobscured, broad-line AGN, which typically vary by only a few percent over such timescales  \citep{MacLeod2012a,VanVelzen2019}.
The double-peaked emission feature around 4680\,\AA\, seen in the post-peak Keck/LRIS spectrum classifies this nuclear event as a BFF in a galaxy hosting an AGN.  
In this section, we discuss our key findings regarding AT\,2021loi in the context of other types of BF nuclear transients.

\subsection{AT\,2021loi as a BFF}
\label{subsec:diss_bffs}

\begin{figure*}
    \includegraphics[width=0.9\textwidth]{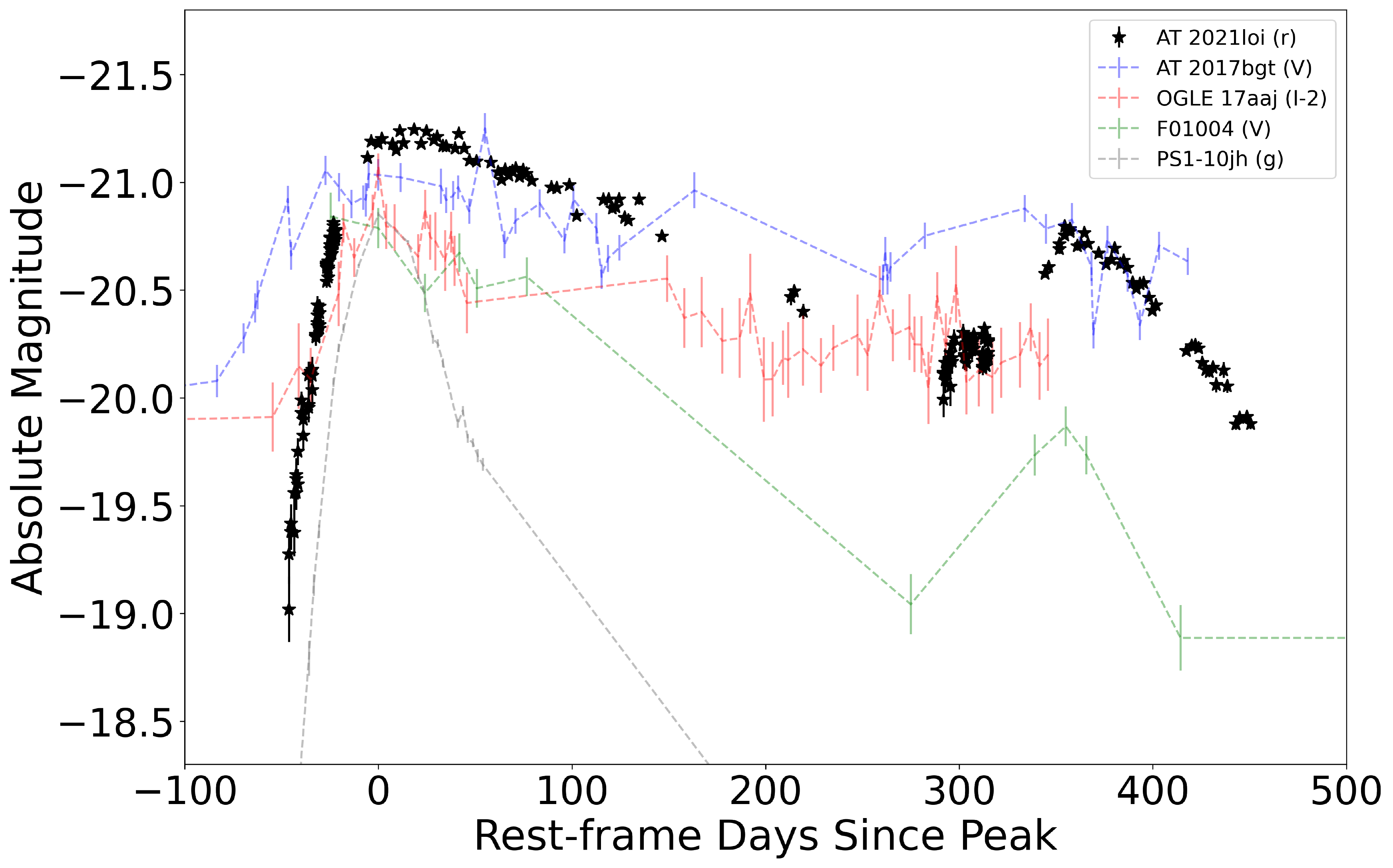}
    \caption{Comparison of the lightcurve of AT\,2021loi with those of the BFFs listed in \citep[][]{Tadhunter2017, Trakhtenbrot2019a,Gromadzki2019} and the TDE PS1-10jh \citep[][]{Gezari2012,Gezari2015}. All the magnitudes are reference subtracted. From the lightcurves, we find thatz although AT\,2021loi has a fast rise, it shows a very slow decline similar to other BFFs and unlike a typical TDE.}
    \label{fig:lccomp}
\end{figure*}

\begin{figure}
    \includegraphics[width=0.45\textwidth]{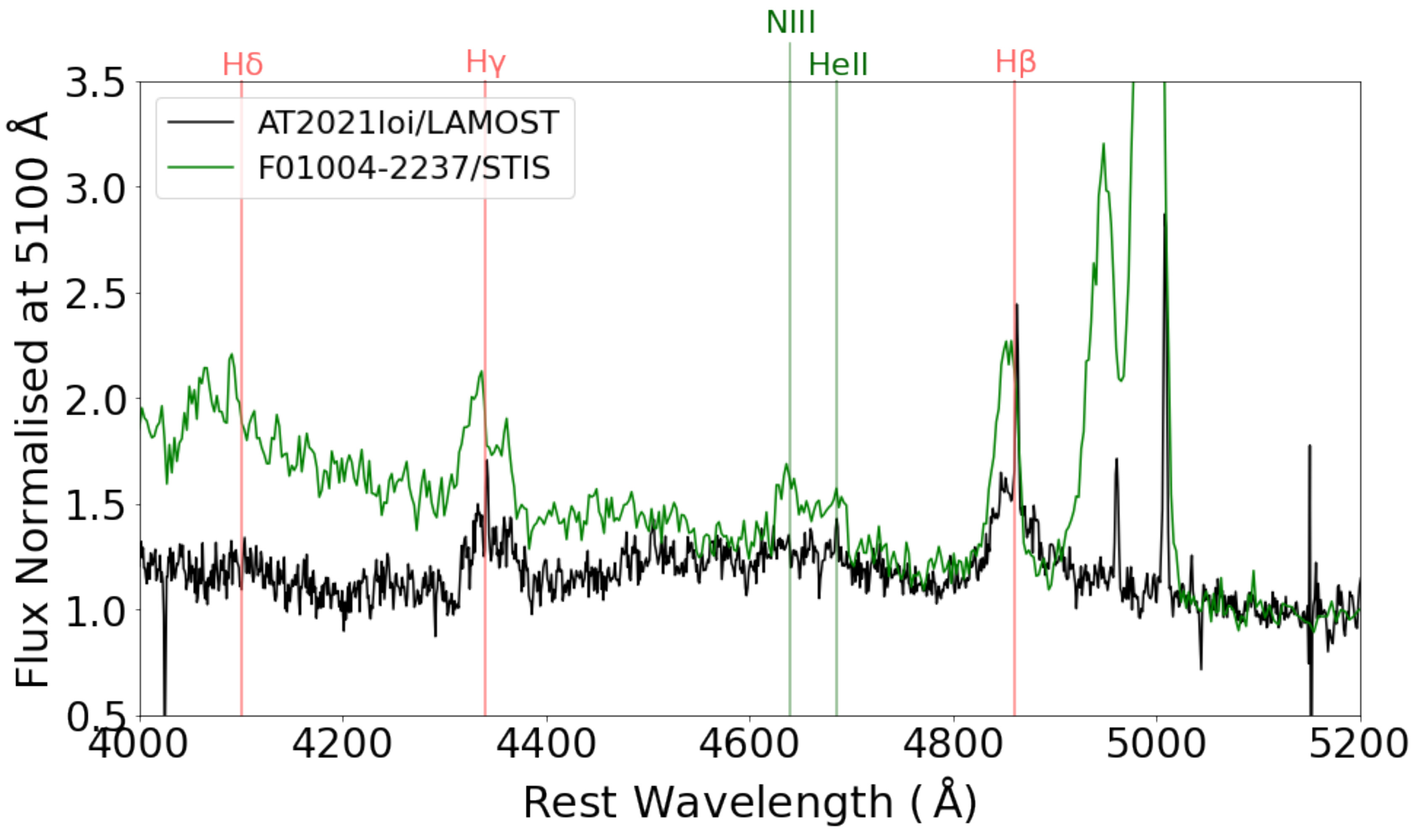}
    \caption{Comparison of the pre-flare LAMOST spectrum of AT\,2021loi with the pre-flare \hst\ spectrum of F01004-2237 \citep{Tadhunter2017}. Both show pre-flare emission features around 4680\AA.}
    \label{fig:loif01004}
\end{figure}

Here we refer to the basic properties of AT\,2021loi that suggest it is part of the BFF class, and compare it to previously reported BFFs. We focus on the three events used by \cite{Trakhtenbrot2019a} to identify this class, namely: AT\,2017bgt (first reported in \citealt{Trakhtenbrot2019a}), F01007-2237 \citep{Tadhunter2017} and OGLE17aaj \citep{Gromadzki2019}.

After the initial increase in brightness, AT\,2021loi shows a very slow decline until the rebrightening $\sim13$ months after the initial detection (see Figures \ref{fig:ltlc}, \ref{fig:lc} and Section \ref{subsec:photev}). The slow decline is comparable to that of the BFFs presented in \cite{Trakhtenbrot2019a} i.e., after more than a year, the optical transient emission still persists. In Figure \ref{fig:lccomp}, we present the $r$-band lightcurve of AT\,2021loi along with those of the \cite{Trakhtenbrot2019a} BFFs. 
The rebrightening and second peak are very clear for AT\,2021loi. 
There is evidence for a rebrightening in F01004-2237, too, which we point out explicitly here for the first time. However, in this case the photometric cadence is rather low and the host subtraction is more challenging. 
The most up-to-date ZTF and ATLAS photometry for AT\,2017bgt is also indicative of another bump, roughly 14 months after the initial detection (and thirteen months after the first peak), however this possible rebrightening is even less robust than the one in F01004-2237.\footnote{The data showing this secondary peak of AT\,2017bgt was not available during the analysis of \cite{Trakhtenbrot2019a}.} 
A late-time rebrightening of the (optical) emission may thus emerge as another intriguing common property of BFFs, although the evidence in hand is still limited.

The optical spectroscopy of AT\,2021loi shows a prominent double-peak emission feature around 4680 \AA, consistent with blended \HeIIop\ and \NIII\ emission. The spectra also show other BF features such as O\,{\sc iii} lines at 3133 and 3444\,\AA. A strong feature around 4680\,\AA\ is also seen in the pre-flare spectrum of AT\,2021loi. To our knowledge, only one of the three BFFs studied by \cite{Trakhtenbrot2019a} has pre-flare optical spectroscopy -- the {\hst}/STIS spectrum of F01004-2237 \citep{Tadhunter2017}. That spectrum also shows pre-existing \NIII\ and \HeIIop\ features, that became stronger following the optical flare. \cite{Tadhunter2017} attributed the pre-existing \NIII\ and \HeIIop\ BF emission to WN Wolf-Rayet stars in the host galaxy, interpreting the flare and \HeIIop\ enhancement as a result of a TDE in F01004-2237.    

The pre-flare $F(\HeIIop$+$\NIII)/F(\hb)$ ratio of AT\,2021loi ($0.58$) is already higher than what is expected for AGN ($\sim0.05$; see, e.g., \citealt{VandenBerk2001}) and is similar to that measured for F01004-2237 ($0.49$, pre-flare). For AT\,2021loi, this ratio increases to $\sim1.1$ near the optical peak, which is slightly lower than what was found in the spectrum of F01004-2237 during its flare ($\sim$1.8) and slightly higher than what was found for OGLE17aaj ($\sim0.9$; \citealt{Gromadzki2019}).   

\begin{figure}
    \includegraphics[width=0.45\textwidth]{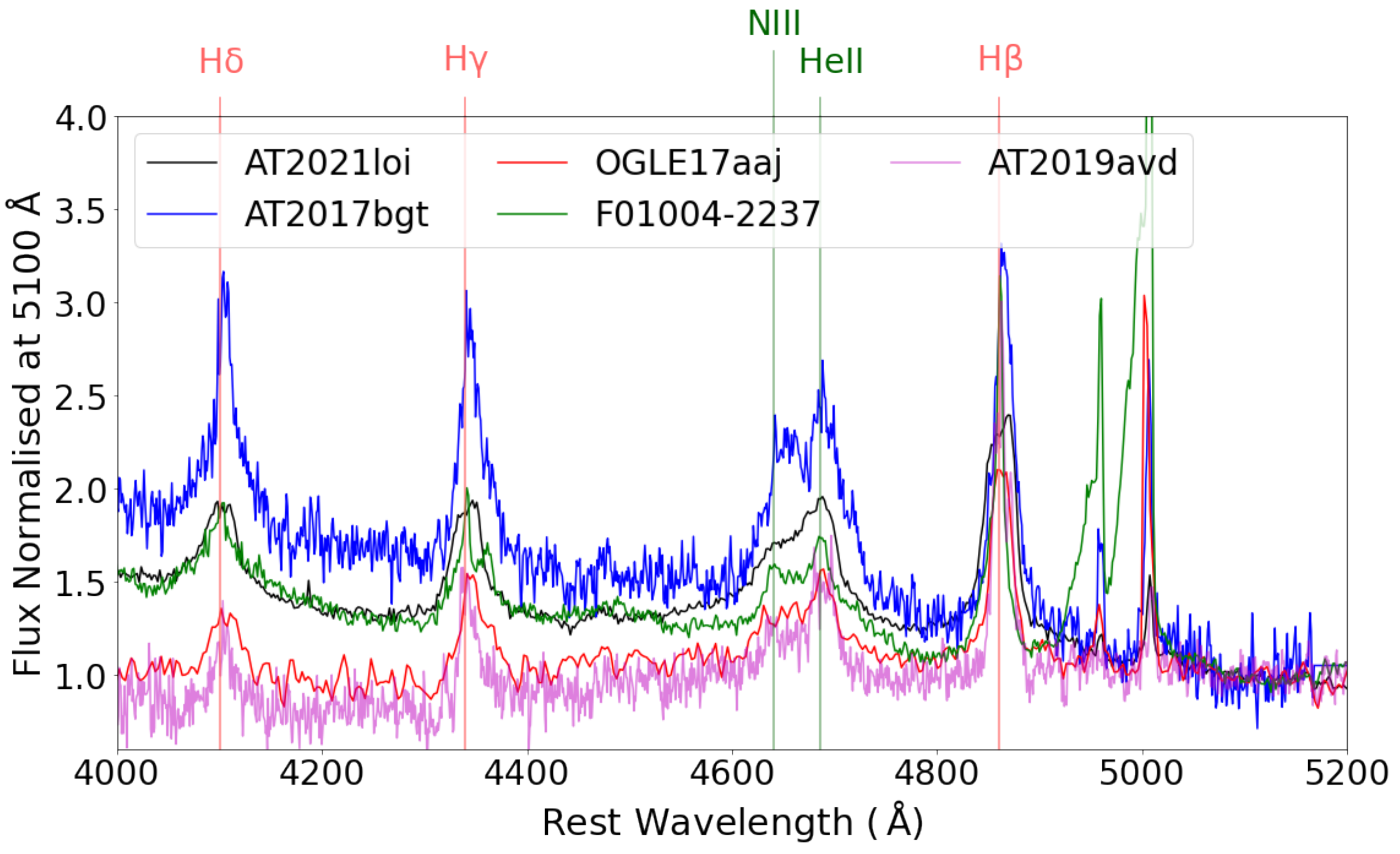}
    \caption{Comparison of the AT\,2021loi spectrum (in blue), taken a month after discovery, with the optical spectra of the three BFFs discussed in \cite{Trakhtenbrot2019a}.}
    \label{fig:bffs}
\end{figure}

In Figure~\ref{fig:bffs} we compare the near-peak spectrum of AT\,2021loi with those of the BFFs discussed in \cite{Trakhtenbrot2019a}. In addition to the $\sim$4680\AA\ feature, AT\,2021loi also shows a broad and possibly double-peaked feature near \Hdelta, similar to what is seen for OGLE17aaj \citep{Gromadzki2019}. As previously mentioned, the \Hdelta\, emission line coincides with the \niii\ BF features expected at 4097 and 4104\,\AA. These two lines are expected to have relative intensities to \NIII\ of 0.117 and 0.082 (respectively; \citealt{Netzer1985}). Given the limited resolution of the AT\,2021loi spectra in hand, it is not possible to properly decompose these lines from \Hdelta\, but we suggest that since the $F(\Hdelta)/F(\hb)$ ratio increased more from pre- to during-flare spectra than the $F(\Hgamma)/F(\hb)$ ratio, the \Hdelta\ spectral feature is affected by intensified N\,{\sc iii} emission. 

AT\,2021loi also shows signs of enhanced emission (between pre-flare and during-flare spectra) in some higher ionization, coronal lines, specifically Fe\,\textsc{x}\,$\lambda$6375. This line was also found in the BF nuclear transient AT\,2019avd \citep{Malyali2021}. In that case, the presence of coronal emission lines was argued to require, and be associated with, the intense soft X-ray radiation seen in that event. The initial discovery of the X-ray flare by e-ROSITA measured an $0.2-2\,\kev$ luminosity of $2.7\times10^{42}\,\ergs$, which is 90 times brighter than their pre-flare upper limits. The X-ray emission of AT\,2019avd remained rather steady for four months, after which it further increased ($L(0.2-2\,\kev)=1.9\times10^{43}\,\ergs$). For AT\,2021loi, there is no X-ray emission detected by our \swift/XRT monitoring, with $3\sigma$ upper limits in the range $(1.5-5.8)\times10^{42}$\,\ergs at 2--10\,\kev. 
While there is no X-ray emission detected, we note that the \emph{expected} X-ray luminosity for AT\,2021loi at 2\kev, based on the NUV flux ($\nuLnu=\sim2.2\times10^{43}\,\ergs$, see Section \ref{subsec:photev}), is higher than what was detected in AT\,2017bgt ($L(2-10\,\kev)=1.2\times10^{43}\,\ergs$) and in AT\,2019avd ($2.7\times10^{42}\,\ergs$).

Thus, the X-ray emission from AT\,2021loi and from the AGN in its pre-flare state is either intrinsically weaker than what is seen in normal, persistently accreting AGN (taking into account the $\approx$0.6 dex uncertainty on the expected X-ray emission; see above); is considerably weaker than what is seen in other BFFs; and/or is delayed compared with what is seen in other BFFs.
It is highly unlikely that intrinisically luminous X-ray radiation is obscured by (dusty) gas along our line-of-sight, given the prominent blue continuum and broad line emission. 
Finally, we cannot rule out a more complex scenario in which the X-ray emitting corona was disrupted by the UV flare, as suggested in at least one other AGN-related UV-luminous transient \citep{Ricci2020}.
At any rate, the X-ray non-detection following the UV-luminous flare in AT\,2021loi does not follow what is seen in normal AGNs.

AT\,2021loi is associated with increased UV emission (by a factor of $\sim20$) compared to the archival GALEX measurement of its host. The pre-flare optical spectrum clearly shows the presence of an actively accreting SMBH in the nucleus of the host galaxy. The rise of the UV emission, peaking at $\nuLnu\simeq2.3\times 10^{44}\ergs$, combined with the simultaneous and significant brightening of the Bowen features, strongly suggests that the AT\,2021loi BF features are driven by the accretion flow onto the SMBH. 

We can use the observed and modeled SEDs of AT\,2021loi to assess whether there is indeed sufficient EUV emission to account for the  BF features.
First, by integrating the best-fit ($17.2\times 10^3$ K) blackbody SED of the flare near peak, we find that only $\sim10^{-12}$ of the total luminosity is emitted below 228\AA\ (the range required to ionize $\textrm{He}\,\textsc{ii}$), corresponding to a luminosity of the order of $10^{32}\,\ergs$. The observed Bowen lines near peak, however, have respective (\HeIIop\ and \NIII) luminosities of order $10^{41}\,\ergs$.
Thus, another ionizing source, outside the measured blackbody continuum, might be required to account for the BF emission. 
This is similar to what was found for the BF-emitting TDE AT\,2018dyb  \citep{Leloudas2019}. 

Second, we can use a (simplified) AGN SED to assess the radiative (EUV) energetics, scaling to either the pre- or during-flare measured luminosities of AT\,2021loi. 
We use the \cite{Marconi2004} SED, scaled to either $\log(\Lbol/\ergs)=43.48$ or $44.46$ for the pre-flare or the during-flare states (respectively). 
The integrated EUV luminosities, i.e. over $\lambda<228$\,\AA, are $7.4\times10^{42}$ and  $5.3\times10^{43}\,\ergs$, for the pre- and during-flare, near peak states (respectively). 
At face value, it appears that the EUV emission from the AGN is  sufficient to account for the energetics of the BF features ($\gtrsim10^{41}\,\ergs$), even before the optical/UV flare. 
However, given that the (optical) luminosity of the pre-existing AGN in this system is not exceptionally high, we recall that BF lines are \emph{not} ubiquitous in persistent AGN \cite[e.g.,][]{Netzer1985,VandenBerk2001}, suggesting that other considerations apart from pure (radiative, EUV) energetics are at play. 

\subsection{Comparison to TDEs}
\label{subsec:diss_tdes}

We next compare AT\,2021loi to the growing sample of TDEs detected in the optical/UV regime (see \citealt{VV2020} and \citealt{Gezari2021_rev} for recent reviews).
The rise time of AT\,2021loi, of about a month, is similar to what is observed in optical/UV TDEs. However, the decline of the optical emission in AT\,2021loi, which can be described by a power law with index shallower than $\alpha\simeq-0.8$ (second decline $\alpha\simeq-0.9$), is much slower than what is seen in optical/UV TDEs, which are typically compared to $\alpha\simeq -5/3$. This is also apparent in Figure \ref{fig:lccomp}, where we compare the optical/UV lightcurve of AT\,2021loi with that of the TDE PS1-10jh \citep{Gezari2012,Gezari2015}. The TDE lightcurve fades within a few months, whereas AT\,2021loi lasts for more than one year, and also presents a secondary brightening. If we fit the post-peak decline of AT\,2021loi with a $t^{-5/3}$ power-law, the required  disruption time is at $\textrm{MJD} = 59031\pm11$ (i.e., 2020 June--July), which is approximately one year before the initial ZTF detection of the flare. 
While the tidal disruptions of \emph{giant} stars are indeed expected to have longer time scales \citep[e.g.;][]{MacLeod2012b}, this would also imply longer rise times (i.e. months instead of weeks), which is not the case with AT\,2021loi. 
We conclude that the optical/UV lightcurve of AT\,2021loi is markedly different than what is seen in (optical/UV) TDEs. We conclude that the overall shape of the AT\, 2021loi is very different from what we know so far for TDEs.

In terms of spectral properties, in many cases optical/UV TDEs also show broad emission features around 4680\AA\ (associated with \HeIIop\ and \NIII; see \citealt{VV2020} and references therein). However, we first note that the emission lines (such as \heii\ and Balmer) in TDEs (FWHM$\sim 10^4\,\kms$) are much broader than those observed in AT\,2021loi (FWHM$\sim 3000\,\kms$). In terms of broad-band SEDs, the color of AT\,2021loi is redder than what is found for TDEs in the early stages and shows very limited evolution (Figure \ref{fig:colourev}). When modeling the transient emission in AT\,2021loi with a blackbody, we find that the derived temperature is much lower than what is seen in TDEs, whereas its (peak) bolometric luminosity and its inferred blackbody radius are higher than in TDEs (see Section \ref{subsec:photev}). 
These features lend further support to our interpretation that the transient emission in AT\,2021loi is not driven by a TDE -- at least not of the common optical/UV kind -- as emphasized by \cite{Trakhtenbrot2019b} for the BFF class as a whole.

AT\,2021loi is also not compatible with scenarios of partial tidal disruption events \citep[e.g.][]{Zhong2022}, double tidal disruption events \citep[e.g.][]{Mandel2015} or extreme mass-ratio inspirals \citep[EMRIs; e.g.,][]{Sari2019}. The second peak in AT\,2021loi occurred $\sim400$ days after the first peak, which is longer than predicted in the partial TDEs scenarios (i.e., $\lesssim$ 150 days for main sequence stars). On the other hand, in the case of EMRIs, repeated flares are expected to happen on longer timescales (i.e., $\gtsim$ several years). However, more detailed simulations of such events are needed to investigate their possible emission signatures.

\section{Conclusions}
\label{concl}

We presented multi-epoch observations of the nuclear transient AT\,2021loi, whose main observed photometric and spectroscopic characteristics are as follows:

\begin{enumerate}
    \item AT\,2021loi is a transient event in a previously known broad-line AGN (WISEA J010039.62+394230.3). An archival spectrum shows clear broad \ha , \hb\, and \Hgamma\, emission and a bump around 4680\AA\, which suggests that at least \HeIIop\, was present before the flare.
    
    \item AT\,2021loi shows a rapid ($\sim$month) rise of the optical flux by a factor of $\sim 4\times$ and a rise in the UV by a factor of $\sim20\times$ while no previous strong variability is found (Figure \ref{fig:ltlc}). AT\,2021loi shows a very slow decline for thirteen months after the initial detection (about twelve months after the peak). Then the optical/UV lightcurve shows a rebrightening and a second peak about 390 days after the first one. 
    
    \item The spectra during the transient phase show enhanced broad \ha, \hb, \Hgamma\, and \Hdelta\, emission. We also find enhanced $\textrm{He\,\textsc{i}}$ emission at 5875 \AA. 
    
    \item The spectra of the source during the flare also show BF lines such as \NIII\ and \OIIIbf.
    
    \item The flare spectroscopy also reveals  signs for enhanced emission from higher ionization coronal lines such as $\textrm{Fe\,{\sc x}}\lambda6375$. The width of this line is consistent with that of other broad lines (i.e., originating from the BLR), unlike what has been seen in AGN until now.
    
    \item AT\,2021loi shows no X-ray emission down to a limit of $L(2-10\,\kev)<2.5\times10^{42}\ergs$ (3$\sigma$) which is considerably lower (by a factor of $\sim 25$) than what is expected from the UV-to-X-ray scaling relations of normal AGN.
\end{enumerate}

AT\,2021loi adds further insight, but also raises new questions, related to the already complex phenomenology of SMBH-related transients and extreme AGN variability. Continued monitoring of AT\,2021loi could provide insights as to its multi-peak nature, testing models of repeated flaring activity. Any late-time X-ray or radio emission could provide further clues as to the nature of this event, and perhaps the class of BFFs as a whole.

\vspace{0.5in}
L.M., B.T., I.A. and M.C.L. acknowledge support from the European Research Council (ERC) under the European Union's Horizon 2020 research and innovation program (grant agreements 852097 and 950533) and from the Israel Science Foundation (grant numbers 1849/19 and 2752/19).
I.A. is a CIFAR Azrieli Global Scholar in the Gravity and the Extreme Universe Program and acknowledges support from that program, from the United States—Israel Binational Science Foundation (BSF), and from the Israeli Council for Higher Education Alon Fellowship.
C.R. acknowledges support from the Fondecyt Iniciacion grant 11190831 and ANID BASAL project FB210003. A.H. is grateful for the support by the I-Core Program of the Planning and Budgeting Committee and the Israel Science Foundation, and support by ISF grant 647/18. A.H. is grateful for support by the Zelman Cowen Academic Initiatives. This publication was made possible through the support of an LSSTC Catalyst Fellowship to K.A.B funded through Grant 62192 from the John Templeton Foundation to LSST Corporation. The opinions expressed in this publication are those of the author(s) and do not necessarily reflect the views of LSSTC or the John Templeton Foundation.

Guoshoujing Telescope (the Large Sky Area Multi-Object Fiber Spectroscopic Telescope LAMOST) is a National Major Scientific Project built by the Chinese Academy of Sciences. Funding for the project has been provided by the National Development and Reform Commission. LAMOST is operated and managed by the National Astronomical Observatories, Chinese Academy of Sciences.The ZTF forced-photometry service was funded under the Heising-Simons Foundation grant No. 12540303 (PI: Graham). This work makes use of data from the Las Cumbres Observatory global telescope network. The LCO group is supported by NSF grant AST-1911151 and AST-1911225 and NASA Swift grant 80NSSC19k1639. The National Radio Astronomy Observatory is a facility of the National Science Foundation operated under cooperative agreement by Associated Universities, Inc. We acknowledge the staff who operate and run the AMI-LA telescope at Lord’s Bridge, Cambridge, for the AMI-LA radio data. AMI is supported by the Universities of Cambridge and Oxford, and by the European Research Council under grant ERC-2012-StG307215 LODESTONE.

This work made use of data from the Asteroid Terrestrial- impact Last Alert System (ATLAS) project. ATLAS is primarily funded to search for near earth asteroids through NASA grants NN12AR55G, 80NSSC18K0284, and 80NSSC18K1575; by-products of the NEO search include images and catalogs from the survey area. This work was partially funded by Kepler/K2 grant J1944/80NSSC19K0112 and HST GO-15889, and STFC grants ST/T000198/1 and ST/S006109/1. The ATLAS science products have been made possible through the contributions of the University of Hawaii Institute for Astronomy, the Queen’s University Belfast, the Space Telescope Science Institute, the South African Astro- nomical Observatory, and The Millennium Institute of Astrophysics (MAS), Chile. 

This work also made use of the NASA/IPAC Extragalactic Database (NED), which is funded by the National Aeronautics and Space Administration and operated by the California Institute of Technology, and of data, software and web tools obtained from the High Energy Astrophysics Science Archive Research Center (HEASARC), a service of the Astrophysics Science Division at NASA/GSFC and of the Smithsonian Astrophysical Observatory’s High Energy Astrophysics Division.


\software{{\tt AstroPy} \citep{astropy:2013, astropy:2018, astropy:2022}, {\tt Matplotlib} \citep{Hunter:2007}, {\tt NumPy} \citep{harris2020array}, {\tt SciPy} \citep{2020SciPy-NMeth}, {\tt lcogtsnpipe} \citep{Valenti2016}, {\tt PyQSOFit} \citep{Guo2018}, {\tt ASPIRED}  \citep{lam_marco_c_2022_6903357}}
\facilities{Las Cumbres (FLOYDS, Sinistro), Keck (LRIS), LAMOST, WISE, Swift (UVOT and XRT), ZTF, ATLAS, VLA, AMI-LA}

\bibliography{AT2021loi.bib}
\bibliographystyle{aasjournal}

\appendix

\section{Noralized Keck and LAMOST spectra}
\label{app:fig_lamostkeck_norm}

\begin{figure*}
    \centering
    \includegraphics[width=0.45\textwidth]{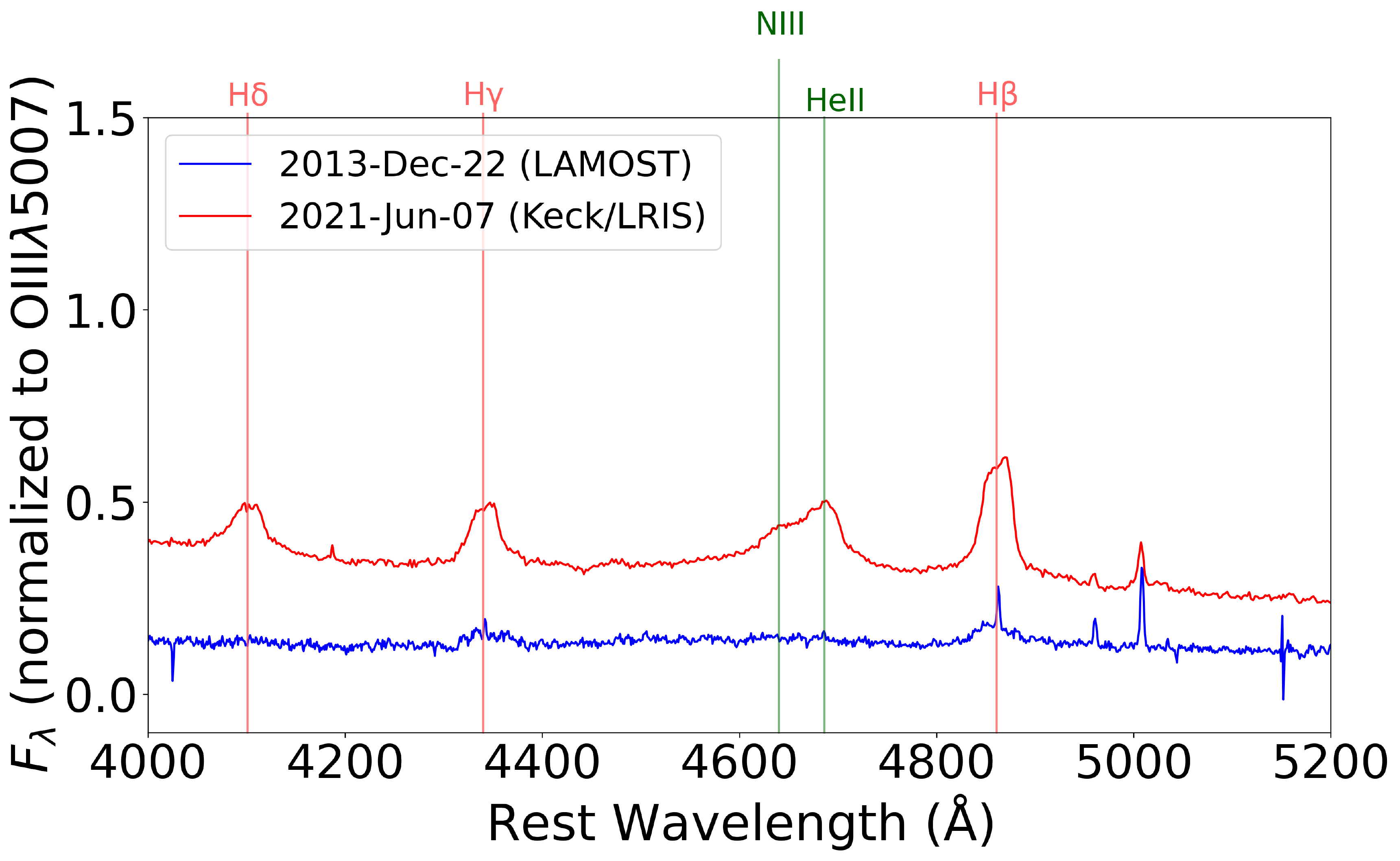}
    \includegraphics[width=0.45\textwidth]{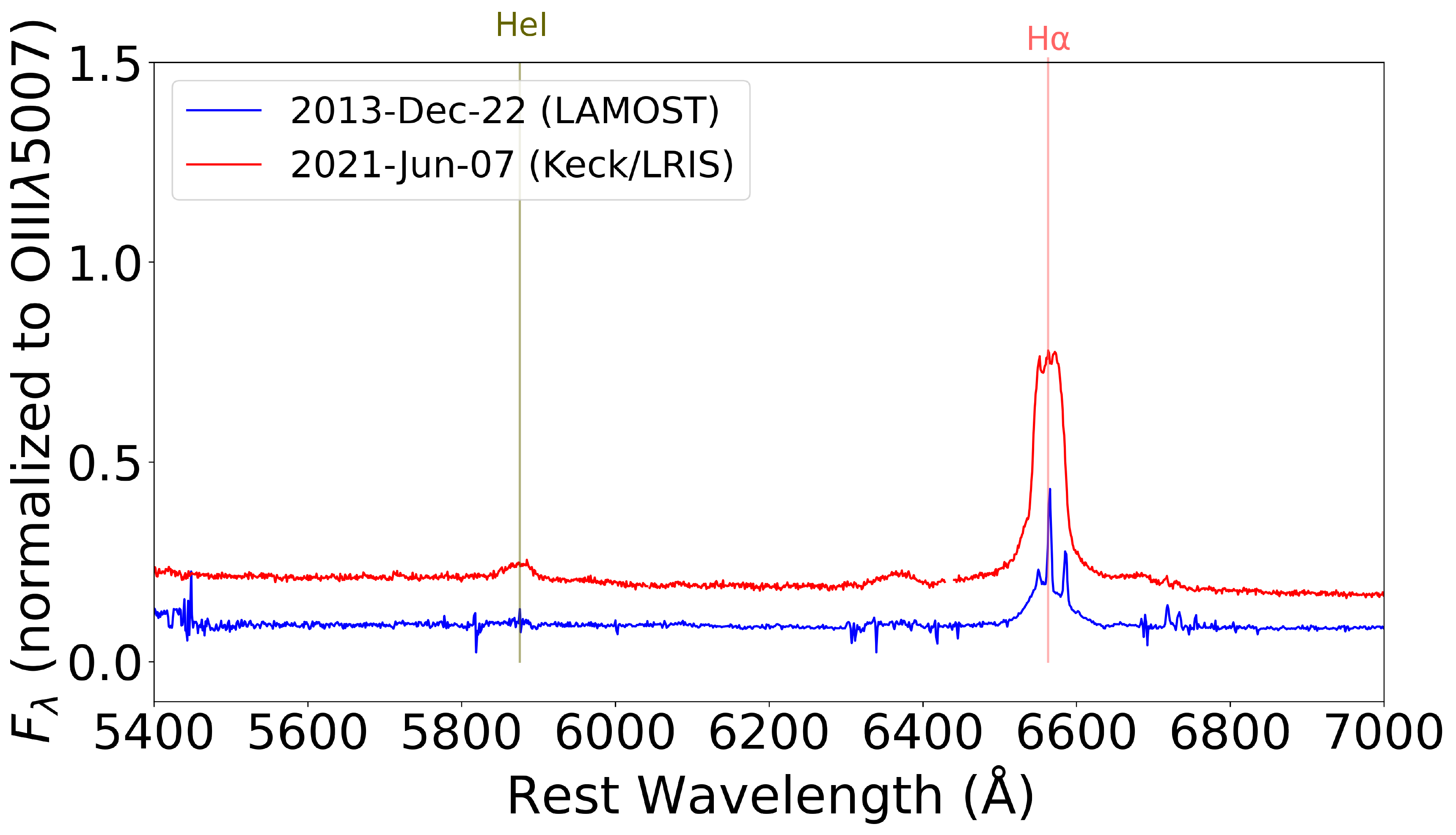}
    \caption{Comparison of the host-subtracted and extinction corrected LAMOST and Keck spectra divided by the \OIII\ flux as found by \textsc{pyqsofit}. 
    } 
    \label{fig:lamostkeck_norm_oiii}
\end{figure*}

Figure~\ref{fig:lamostkeck_norm_oiii} shows the archival LAMOST and near-peak (classification) Keck spectra of AT\,2021loi, but with flux densities normalized such that the integrated flux of the best-fitting \OIII\ line profile would result in 1 ${\rm erg\,s^{-1}\,cm^{-2}}$. 
To derive the corresponding normalization factor for each spectrum, we employed the \textsc{pyqsofit} spectral fitting procedure (see Section~\ref{subsec:specev}), but note that in this case we did not try to assign any physical meaning to the continuum decomposition, broad line profiles, or other parameters.
As we do not expect the \OIII\ to vary within timescales of a few years \cite[see discussion, and a rare counter-example, in][]{Peterson2013}, these normalized spectra serve to further highlight the differences in continuum and broad line emission between the two epochs, independent of potential seeing and aperture effects.

\section{Details regarding optical spectra}
\label{app:tab_opt_spec}

Table~\ref{tab:opt_spec} lists basic details regarding the optical spectra we use and some key spectral measurements, derived through our spectral fitting (see Section~\ref{subsec:specev}). 

\begin{deluxetable}{cccccccc}
\tabletypesize{\footnotesize}
\tablecolumns{8} 
\tablecaption{Optical Spectroscopy of AT\,2021loi} 
\tablehead{\colhead{MJD} & \colhead{Telescope/Instrument} &  \colhead{$R$\tablenotemark{a}} & \colhead{$F(\hb)$} & \colhead{$F({\rm BF})$\tablenotemark{b}} & \colhead{$F(\OIII)$} & \colhead{$\log\Lop$} & \colhead{FWHM(\hb)}\\
\nocolhead{} & \nocolhead{} & \colhead{(approx.)} &  \colhead{$(10^{-15}\,\ergcms$}) & \colhead{$(10^{-15}\,\ergcms)$} & \colhead{$(10^{-15}\,\ergcms)$} & \colhead{(\ergs)} & \colhead{(\kms)} } 
\startdata
56648.00 & LAMOST & 1800 & 10.12 $\pm$ 0.54 & 5.82 $\pm$ 0.46& 2.29 $\pm$ 0.12 & 42.5 & 3024 $\pm$ 248\\
\hline
59372.61 & Keck/LRIS & 900 & 47.12 $\pm$ 2.13&  55.88 $\pm$ 2.96&  2.42 $\pm$ 0.11 & 43.48 & 2146 $\pm$ 31\\ 
59378.59 & Las Cumbres/FLOYDS-N & 500 & 39.41 $\pm$ 2.06 & 35.31 $\pm$ 1.85 & 2.20 $\pm$ 0.12 & 43.59 & 2087 $\pm$ 94\\ 
59399.60 & Las Cumbres/FLOYDS-N & 500  &58.42 $\pm$ 2.55&  40.49 $\pm$ 1.76& 2.48 $\pm$ 0.11& 43.65 & 2092 $\pm$ 61\\ 
59407.55 & Las Cumbres/FLOYDS-N & 500 & 53.16 $\pm$ 2.42&  25.91 $\pm$ 1.18& 2.54 $\pm$ 0.12& 43.51 & 1917 $\pm$ 77\\
59415.53 & Las Cumbres/FLOYDS-N & 500  &63.08 $\pm$ 3.47&  29.82 $\pm$ 1.64&  3.39 $\pm$ 0.19& 43.53 & 2162 $\pm$ 63\\ 
59429.46 & Las Cumbres/FLOYDS-N & 500  &78.90 $\pm$ 2.52&  31.77 $\pm$ 1.02& 4.17 $\pm$ 0.13& 43.65 & 2051 $\pm$ 120\\
59446.52 & Las Cumbres/FLOYDS-N & 500  &69.93 $\pm$ 0.97&   22.13 $\pm$ 0.31&  3.08 $\pm$ 0.04&  43.48 & 2078 $\pm$ 52\\ 
59454.61 & Las Cumbres/FLOYDS-N & 500  &65.78 $\pm$ 1.14&  21.25 $\pm$ 0.37&  2.52 $\pm$ 0.04&  43.44 & 2125 $\pm$ 49\\ 
59462.49 & Las Cumbres/FLOYDS-N & 500  &62.87 $\pm$ 1.27&  19.08 $\pm$ 0.38&  2.78 $\pm$ 0.06&  43.41 & 2184 $\pm$ 80\\ 
59470.44 & Las Cumbres/FLOYDS-N & 500  &72.48 $\pm$ 1.82&  21.76 $\pm$ 0.55&  3.25 $\pm$ 0.08&  43.43 & 2171 $\pm$ 101\\
59478.44 & Las Cumbres/FLOYDS-N & 500 & 69.84 $\pm$ 4.29&  22.58 $\pm$ 1.39& 1.92 $\pm$ 0.12&  43.25 & 2279 $\pm$ 189\\
59489.37 & Las Cumbres/FLOYDS-N & 500  &65.38 $\pm$ 1.72&  23.26 $\pm$ 0.61&  2.58 $\pm$ 0.07&  43.29 & 2409 $\pm$ 100\\ 
59504.28 & Las Cumbres/FLOYDS-N &500  &48.26 $\pm$ 0.88&    12.11 $\pm$ 0.22&   1.04 $\pm$ 0.02&  42.99 & 2272 $\pm$ 53\\ 
59507.37 & Las Cumbres/FLOYDS-N & 500  &60.37 $\pm$ 2.93&  15.40 $\pm$ 0.75&  1.25 $\pm$ 0.06 &  42.63 & 2761 $\pm$ 103\\
59522.34 & Las Cumbres/FLOYDS-N & 500  &52.01 $\pm$ 1.40&   17.02 $\pm$ 0.46&   1.68 $\pm$ 0.05&  42.95 & 2440 $\pm$69\\  
59537.35 & Las Cumbres/FLOYDS-N & 500  &67.34 $\pm$ 1.06&  19.30 $\pm$ 0.30&  2.44 $\pm$ 0.04&  43.18 & 2370 $\pm$ 44\\ 
59585.29 & Las Cumbres/FLOYDS-N & 500  &54.20 $\pm$ 0.98&   17.20 $\pm$ 0.31&  1.88 $\pm$ 0.03&   42.69 & 2365 $\pm$74\\ 
59601.24 & Las Cumbres/FLOYDS-N & 500  &70.13 $\pm$ 1.08&  19.88 $\pm$ 0.31&  2.62 $\pm$ 0.04&  43.11 & 2340 $\pm$ 57\\ 
59616.22 & Las Cumbres/FLOYDS-N & 500  &66.00 $\pm$ 0.98&   17.97 $\pm$ 0.27&  2.62 $\pm$ 0.04& 43.0 & 2265 $\pm$ 54\\ 
59804.55 & Las Cumbres/FLOYDS-N & 500 & 72.00 $\pm$ 0.91&   24.61 $\pm$ 0.31&  2.00 $\pm$ 0.03&  42.46 & 2312 $\pm$ 148\\
59839.52 & Las Cumbres/FLOYDS-N & 500 & 57.93 $\pm$ 9.19& 8.00 $\pm$ 1.27&  1.50 $\pm$ 0.23& 42.63 & 2333 $\pm$ 528
\enddata 
\tablenotetext{a}{The (approximate) resolving power of the spectrum.}
\tablenotetext{b}{With the term BF here, we mean the double peaked emission feature around $\sim$4680\,\AA\ originating from \HeIIop\ and \NIII.}
\label{tab:opt_spec}
\end{deluxetable}



\end{document}
